\pdfoutput=1
\documentclass[twocolumn]{aastex631}

\usepackage{amsmath}
\usepackage{amssymb}
\usepackage{graphicx}
\usepackage{booktabs}

\newcommand{\BridgeJWSTEuclidContinuumGap}{0.218}

\newcommand{\EuclidBridgeCMin}{44.463}

\newcommand{\EuclidBridgeN}{280}

\newcommand{\EuclidHLimit}{20.10}

\newcommand{\EuclidMainLHaMax}{43.86}

\newcommand{\EuclidMainLHaMin}{42.82}
\newcommand{\EuclidMainMedFWHM}{4251}
\newcommand{\EuclidMainN}{334}

\newcommand{\EuclidMainSGScatterUniform}{0.1384}

\newcommand{\EuclidMainSigmaW}{0.143}
\newcommand{\EuclidMainSigmaWLowerNinetyFive}{0.133}

\newcommand{\EuclidMainSigmaWUpperNinetyFive}{0.155}

\newcommand{\EuclidMainZMax}{1.72}
\newcommand{\EuclidMainZMed}{1.02}
\newcommand{\EuclidMainZMin}{0.88}

\newcommand{\EuclidTargets}{914}

\newcommand{\JWSTBridgeCMax}{44.245}

\newcommand{\JWSTBridgeN}{3}
\newcommand{\JWSTCleanAtlasProvN}{22}
\newcommand{\JWSTCleanBXProvN}{30}
\newcommand{\JWSTCleanLHaMax}{43.83}
\newcommand{\JWSTCleanLHaMed}{42.48}
\newcommand{\JWSTCleanLHaMin}{41.69}
\newcommand{\JWSTCleanMedFWHM}{1929}
\newcommand{\JWSTCleanN}{40}
\newcommand{\JWSTCleanSGScatterUniform}{0.0632}

\newcommand{\JWSTCleanWangOverlapN}{27}
\newcommand{\JWSTCleanZMax}{6.99}
\newcommand{\JWSTCleanZMed}{5.28}
\newcommand{\JWSTCleanZMin}{2.98}
\newcommand{\JWSTEdgeFallbackN}{2}
\newcommand{\JWSTGratingParentN}{43}
\newcommand{\JWSTGratingParentWangOverlapN}{30}

\newcommand{\JWSTHighSNRMedFWHM}{2173}
\newcommand{\JWSTHighSNRN}{22}

\newcommand{\JWSTHighSNRWangOverlapN}{15}
\newcommand{\JWSTHighSNRZMax}{6.99}

\newcommand{\JWSTHighSNRZMin}{3.43}

\newcommand{\JWSTOtherCleanN}{20}
\newcommand{\JWSTPresetWangOverlapN}{14}
\newcommand{\JWSTPresetZMax}{6.99}

\newcommand{\JWSTPresetZMin}{3.43}

\newcommand{\JWSTTwoLineInCleanN}{30}
\newcommand{\JWSTTwoLineOutCleanN}{3}
\newcommand{\JWSTTwoLineWangOverlapN}{21}
\newcommand{\JWSTTwoLineZMax}{7.15}
\newcommand{\JWSTTwoLineZMin}{2.98}

\newcommand{\POneASrcFiveFiveOneFiveTwoFiveD}{8.7}
\newcommand{\POneASrcFiveFiveOneFiveTwoFiveR}{0.115}
\newcommand{\POneASrcFiveFiveOneFiveTwoFiveRErr}{0.013}
\newcommand{\POneASrcFiveFiveTenOneFiveD}{11.1}
\newcommand{\POneASrcFiveFiveTenOneFiveR}{0.090}
\newcommand{\POneASrcFiveFiveTenOneFiveRErr}{0.009}
\newcommand{\POneASrcFiveFiveThreeTenD}{12.5}
\newcommand{\POneASrcFiveFiveThreeTenR}{0.080}
\newcommand{\POneASrcFiveFiveThreeTenRErr}{0.005}
\newcommand{\POneASrcFiveFiveTwoFiveThreeFiveD}{5.7}
\newcommand{\POneASrcFiveFiveTwoFiveThreeFiveR}{0.177}
\newcommand{\POneASrcFiveFiveTwoFiveThreeFiveRErr}{0.036}
\newcommand{\POneASrcFiveFiveZeroThreeD}{16.6}
\newcommand{\POneASrcFiveFiveZeroThreeR}{0.060}
\newcommand{\POneASrcFiveFiveZeroThreeRErr}{0.005}
\newcommand{\POneASrcFourTwoOneFiveTwoFiveD}{9.5}
\newcommand{\POneASrcFourTwoOneFiveTwoFiveR}{0.105}
\newcommand{\POneASrcFourTwoOneFiveTwoFiveRErr}{0.018}
\newcommand{\POneASrcFourTwoTenOneFiveD}{9.4}
\newcommand{\POneASrcFourTwoTenOneFiveR}{0.107}
\newcommand{\POneASrcFourTwoTenOneFiveRErr}{0.014}
\newcommand{\POneASrcFourTwoThreeTenBlueD}{15.6}
\newcommand{\POneASrcFourTwoThreeTenD}{12.2}
\newcommand{\POneASrcFourTwoThreeTenR}{0.082}
\newcommand{\POneASrcFourTwoThreeTenRErr}{0.007}
\newcommand{\POneASrcFourTwoThreeTenRedD}{11.1}
\newcommand{\POneASrcFourTwoTwoFiveThreeFiveR}{0.063}
\newcommand{\POneASrcFourTwoTwoFiveThreeFiveRErr}{0.048}
\newcommand{\POneASrcFourTwoZeroThreeD}{19.2}
\newcommand{\POneASrcFourTwoZeroThreeR}{0.052}
\newcommand{\POneASrcFourTwoZeroThreeRErr}{0.010}
\newcommand{\POneASrcSixEightOneFiveTwoFiveD}{12.0}
\newcommand{\POneASrcSixEightOneFiveTwoFiveR}{0.084}
\newcommand{\POneASrcSixEightOneFiveTwoFiveRErr}{0.025}
\newcommand{\POneASrcSixEightTenOneFiveD}{19.0}
\newcommand{\POneASrcSixEightTenOneFiveR}{0.052}
\newcommand{\POneASrcSixEightTenOneFiveRErr}{0.021}
\newcommand{\POneASrcSixEightThreeTenD}{16.5}
\newcommand{\POneASrcSixEightThreeTenR}{0.061}
\newcommand{\POneASrcSixEightThreeTenRErr}{0.010}
\newcommand{\POneASrcSixEightTwoFiveThreeFiveR}{0.098}
\newcommand{\POneASrcSixEightTwoFiveThreeFiveRErr}{0.070}
\newcommand{\POneASrcSixEightZeroThreeD}{23.5}
\newcommand{\POneASrcSixEightZeroThreeR}{0.043}
\newcommand{\POneASrcSixEightZeroThreeRErr}{0.012}
\newcommand{\POneBSrcFiveFiveOneFiveTwoFiveD}{9.7}
\newcommand{\POneBSrcFiveFiveOneFiveTwoFiveR}{0.103}
\newcommand{\POneBSrcFiveFiveOneFiveTwoFiveRErr}{0.014}
\newcommand{\POneBSrcFiveFiveTenOneFiveD}{11.9}
\newcommand{\POneBSrcFiveFiveTenOneFiveR}{0.084}
\newcommand{\POneBSrcFiveFiveTenOneFiveRErr}{0.009}
\newcommand{\POneBSrcFiveFiveThreeTenD}{13.1}
\newcommand{\POneBSrcFiveFiveThreeTenR}{0.076}
\newcommand{\POneBSrcFiveFiveThreeTenRErr}{0.005}
\newcommand{\POneBSrcFiveFiveTwoFiveThreeFiveD}{7.1}
\newcommand{\POneBSrcFiveFiveTwoFiveThreeFiveR}{0.140}
\newcommand{\POneBSrcFiveFiveTwoFiveThreeFiveRErr}{0.035}
\newcommand{\POneBSrcFiveFiveZeroThreeD}{14.2}
\newcommand{\POneBSrcFiveFiveZeroThreeR}{0.070}
\newcommand{\POneBSrcFiveFiveZeroThreeRErr}{0.004}
\newcommand{\POneBSrcFourTwoOneFiveTwoFiveD}{8.7}
\newcommand{\POneBSrcFourTwoOneFiveTwoFiveR}{0.115}
\newcommand{\POneBSrcFourTwoOneFiveTwoFiveRErr}{0.020}
\newcommand{\POneBSrcFourTwoTenOneFiveD}{9.4}
\newcommand{\POneBSrcFourTwoTenOneFiveR}{0.106}
\newcommand{\POneBSrcFourTwoTenOneFiveRErr}{0.014}
\newcommand{\POneBSrcFourTwoThreeTenD}{12.1}
\newcommand{\POneBSrcFourTwoThreeTenR}{0.083}
\newcommand{\POneBSrcFourTwoThreeTenRErr}{0.007}
\newcommand{\POneBSrcFourTwoTwoFiveThreeFiveR}{0.090}
\newcommand{\POneBSrcFourTwoTwoFiveThreeFiveRErr}{0.052}
\newcommand{\POneBSrcFourTwoZeroThreeD}{11.5}
\newcommand{\POneBSrcFourTwoZeroThreeR}{0.087}
\newcommand{\POneBSrcFourTwoZeroThreeRErr}{0.006}
\newcommand{\POneBSrcSixEightOneFiveTwoFiveR}{0.038}
\newcommand{\POneBSrcSixEightOneFiveTwoFiveRErr}{0.028}
\newcommand{\POneBSrcSixEightTenOneFiveR}{0.027}
\newcommand{\POneBSrcSixEightTenOneFiveRErr}{0.021}
\newcommand{\POneBSrcSixEightThreeTenD}{22.3}
\newcommand{\POneBSrcSixEightThreeTenR}{0.045}
\newcommand{\POneBSrcSixEightThreeTenRErr}{0.010}
\newcommand{\POneBSrcSixEightTwoFiveThreeFiveR}{-0.022}
\newcommand{\POneBSrcSixEightTwoFiveThreeFiveRErr}{0.076}
\newcommand{\POneBSrcSixEightZeroThreeD}{11.3}
\newcommand{\POneBSrcSixEightZeroThreeR}{0.089}
\newcommand{\POneBSrcSixEightZeroThreeRErr}{0.009}

\newcommand{\POneSrcFiveFiveNCombine}{3}
\newcommand{\POneSrcFiveFiveZOIII}{6.9834}

\newcommand{\POneSrcFourTwoNCombine}{6}
\newcommand{\POneSrcFourTwoZOIII}{5.2763}

\newcommand{\POneSrcSixEightNCombine}{3}
\newcommand{\POneSrcSixEightZOIII}{5.0391}
\newcommand{\POneTmplAllThirtyThreeHi}{1.274}
\newcommand{\POneTmplAllThirtyThreeLo}{0.780}
\newcommand{\POneTmplAllThirtyThreeN}{33}
\newcommand{\POneTmplAllThirtyThreeRatio}{1.009}

\newcommand{\POneTmplOtherThirtyHi}{1.953}
\newcommand{\POneTmplOtherThirtyLo}{-0.073}
\newcommand{\POneTmplOtherThirtyN}{30}
\newcommand{\POneTmplOtherThirtyRatio}{0.802}
\newcommand{\POneTmplOtherThirtyWtPct}{12.5}
\newcommand{\POneTmplTopThreeHi}{1.311}
\newcommand{\POneTmplTopThreeLo}{0.800}
\newcommand{\POneTmplTopThreeN}{3}
\newcommand{\POneTmplTopThreeRatio}{1.031}
\newcommand{\POneTmplTopThreeWtPct}{87.5}
\newcommand{\PTCalJWSTAttemptedTargetsN}{22}

\newcommand{\PTCalJWSTFortyGaussFourFiveErrSixEight}{0.116}
\newcommand{\PTCalJWSTFortyGaussFourFiveUsablePct}{81.0}

\newcommand{\PTCalJWSTGaussFiveFiveBias}{+0.006}
\newcommand{\PTCalJWSTGaussFiveFiveErrSixEight}{0.086}
\newcommand{\PTCalJWSTGaussFiveFiveUsablePct}{81.4}
\newcommand{\PTCalJWSTGaussFourFiveBias}{+0.001}
\newcommand{\PTCalJWSTGaussFourFiveErrSixEight}{0.057}
\newcommand{\PTCalJWSTGaussFourFiveUsablePct}{86.8}
\newcommand{\PTCalJWSTGaussOneFiveBias}{-0.002}
\newcommand{\PTCalJWSTGaussOneFiveErrSixEight}{0.008}
\newcommand{\PTCalJWSTGaussOneFiveOkTrials}{200}
\newcommand{\PTCalJWSTGaussOneFiveTotTrials}{220}
\newcommand{\PTCalJWSTGaussOneFiveUsablePct}{90.9}
\newcommand{\PTCalJWSTGaussSixZeroBias}{+0.010}
\newcommand{\PTCalJWSTGaussSixZeroErrSixEight}{0.089}
\newcommand{\PTCalJWSTGaussSixZeroUsablePct}{77.3}
\newcommand{\PTCalJWSTGaussThreeFiveBias}{0.000}
\newcommand{\PTCalJWSTGaussThreeFiveErrSixEight}{0.031}
\newcommand{\PTCalJWSTGaussThreeFiveUsablePct}{90.5}
\newcommand{\PTCalJWSTGaussTwoFiveBias}{-0.002}
\newcommand{\PTCalJWSTGaussTwoFiveErrSixEight}{0.013}
\newcommand{\PTCalJWSTGaussTwoFiveUsablePct}{90.9}
\newcommand{\PTCalJWSTWingFiveFiveBias}{-0.012}
\newcommand{\PTCalJWSTWingFiveFiveErrSixEight}{0.056}
\newcommand{\PTCalJWSTWingFiveFiveUsablePct}{85.5}
\newcommand{\PTCalJWSTWingFourFiveBias}{-0.009}
\newcommand{\PTCalJWSTWingFourFiveErrSixEight}{0.038}
\newcommand{\PTCalJWSTWingFourFiveUsablePct}{90.5}
\newcommand{\PTCalJWSTWingOneFiveBias}{-0.002}
\newcommand{\PTCalJWSTWingOneFiveErrSixEight}{0.009}
\newcommand{\PTCalJWSTWingOneFiveUsablePct}{90.9}
\newcommand{\PTCalJWSTWingSixZeroBias}{-0.019}
\newcommand{\PTCalJWSTWingSixZeroErrSixEight}{0.064}
\newcommand{\PTCalJWSTWingSixZeroUsablePct}{84.5}
\newcommand{\PTCalJWSTWingThreeFiveBias}{-0.005}
\newcommand{\PTCalJWSTWingThreeFiveErrSixEight}{0.024}
\newcommand{\PTCalJWSTWingThreeFiveUsablePct}{90.9}
\newcommand{\PTCalJWSTWingTwoFiveBias}{-0.002}
\newcommand{\PTCalJWSTWingTwoFiveErrSixEight}{0.013}
\newcommand{\PTCalJWSTWingTwoFiveUsablePct}{90.9}
\newcommand{\PTCalSDSSAttemptedTargetsN}{185}
\newcommand{\PTCalSDSSGaussFiveFiveBias}{-0.002}
\newcommand{\PTCalSDSSGaussFiveFiveErrSixEight}{0.023}
\newcommand{\PTCalSDSSGaussFiveFiveUsablePct}{90.3}
\newcommand{\PTCalSDSSGaussFourFiveBias}{-0.002}
\newcommand{\PTCalSDSSGaussFourFiveErrSixEight}{0.014}
\newcommand{\PTCalSDSSGaussFourFiveUsablePct}{90.8}
\newcommand{\PTCalSDSSGaussOneFiveBias}{-0.001}
\newcommand{\PTCalSDSSGaussOneFiveErrSixEight}{0.005}
\newcommand{\PTCalSDSSGaussOneFiveOkTrials}{1700}
\newcommand{\PTCalSDSSGaussOneFiveTotTrials}{1850}
\newcommand{\PTCalSDSSGaussOneFiveUsablePct}{91.9}
\newcommand{\PTCalSDSSGaussSixZeroBias}{-0.002}
\newcommand{\PTCalSDSSGaussSixZeroErrSixEight}{0.030}
\newcommand{\PTCalSDSSGaussSixZeroUsablePct}{89.7}
\newcommand{\PTCalSDSSGaussThreeFiveBias}{-0.001}
\newcommand{\PTCalSDSSGaussThreeFiveErrSixEight}{0.008}
\newcommand{\PTCalSDSSGaussThreeFiveUsablePct}{91.4}
\newcommand{\PTCalSDSSGaussTwoFiveBias}{-0.001}
\newcommand{\PTCalSDSSGaussTwoFiveErrSixEight}{0.005}
\newcommand{\PTCalSDSSGaussTwoFiveUsablePct}{91.9}
\newcommand{\PTCalSDSSWingFiveFiveBias}{-0.014}
\newcommand{\PTCalSDSSWingFiveFiveErrSixEight}{0.024}
\newcommand{\PTCalSDSSWingFiveFiveUsablePct}{90.5}
\newcommand{\PTCalSDSSWingFourFiveBias}{-0.008}
\newcommand{\PTCalSDSSWingFourFiveErrSixEight}{0.015}
\newcommand{\PTCalSDSSWingFourFiveUsablePct}{90.7}
\newcommand{\PTCalSDSSWingOneFiveBias}{-0.002}
\newcommand{\PTCalSDSSWingOneFiveErrSixEight}{0.005}
\newcommand{\PTCalSDSSWingOneFiveUsablePct}{91.8}
\newcommand{\PTCalSDSSWingSixZeroBias}{-0.018}
\newcommand{\PTCalSDSSWingSixZeroErrSixEight}{0.028}
\newcommand{\PTCalSDSSWingSixZeroUsablePct}{90.5}
\newcommand{\PTCalSDSSWingThreeFiveBias}{-0.004}
\newcommand{\PTCalSDSSWingThreeFiveErrSixEight}{0.010}
\newcommand{\PTCalSDSSWingThreeFiveUsablePct}{91.3}
\newcommand{\PTCalSDSSWingTwoFiveBias}{-0.002}
\newcommand{\PTCalSDSSWingTwoFiveErrSixEight}{0.006}
\newcommand{\PTCalSDSSWingTwoFiveUsablePct}{91.8}
\newcommand{\PTDiffFourFive}{+0.0069}
\newcommand{\PTDiffHiFourFive}{+0.0269}
\newcommand{\PTDiffHiThreeFive}{-0.0017}
\newcommand{\PTDiffHiTwoFive}{-0.0047}
\newcommand{\PTDiffLoFourFive}{-0.0221}
\newcommand{\PTDiffLoThreeFive}{-0.0316}
\newcommand{\PTDiffLoTwoFive}{-0.0274}
\newcommand{\PTDiffThreeFive}{-0.0144}
\newcommand{\PTDiffTwoFive}{-0.0144}
\newcommand{\PTEdgeFiveZeroZeroFiveTwoGaussErrSixEight}{0.07}
\newcommand{\PTEdgeFiveZeroZeroFiveTwoGaussUsablePct}{91.7}
\newcommand{\PTEdgeFiveZeroZeroFiveTwoWingErrSixEight}{0.07}
\newcommand{\PTEdgeFiveZeroZeroFiveTwoWingUsablePct}{100.0}
\newcommand{\PTEdgeThreeEightFiveSixTwoGaussErrSixEight}{0.17}
\newcommand{\PTEdgeThreeEightFiveSixTwoGaussUsablePct}{63.3}
\newcommand{\PTEdgeThreeEightFiveSixTwoWingErrSixEight}{0.14}
\newcommand{\PTEdgeThreeEightFiveSixTwoWingUsablePct}{85.0}
\newcommand{\PTJWSTFortyNFourFive}{38}
\newcommand{\PTJWSTFortySFourFive}{0.1417}
\newcommand{\PTJWSTFortySTwoFive}{0.0739}

\newcommand{\PTJWSTFullDeltaHi}{0.0589}
\newcommand{\PTJWSTFullDeltaLo}{0.0214}
\newcommand{\PTJWSTFullDeltaSigned}{+0.0428}

\newcommand{\PTJWSTFullN}{22}
\newcommand{\PTJWSTFullSFourFive}{0.0811}

\newcommand{\PTJWSTFullSHiFourFive}{0.0992}
\newcommand{\PTJWSTFullSHiOneFive}{0.0383}
\newcommand{\PTJWSTFullSHiThreeFive}{0.0649}
\newcommand{\PTJWSTFullSHiTwoFive}{0.0467}
\newcommand{\PTJWSTFullSLoFourFive}{0.0528}
\newcommand{\PTJWSTFullSLoOneFive}{0.0203}
\newcommand{\PTJWSTFullSLoThreeFive}{0.0355}
\newcommand{\PTJWSTFullSLoTwoFive}{0.0259}
\newcommand{\PTJWSTFullSOneFive}{0.0309}
\newcommand{\PTJWSTFullSThreeFive}{0.0530}
\newcommand{\PTJWSTFullSTwoFive}{0.0383}

\newcommand{\PTJWSTPresetDeltaHi}{0.0643}
\newcommand{\PTJWSTPresetDeltaHiThreeDec}{0.064}
\newcommand{\PTJWSTPresetDeltaLo}{0.0259}
\newcommand{\PTJWSTPresetDeltaLoThreeDec}{0.026}
\newcommand{\PTJWSTPresetDeltaSigned}{+0.0479}

\newcommand{\PTJWSTPresetDeltaThreeDec}{0.048}

\newcommand{\PTJWSTPresetMedSigFourFive}{1225}

\newcommand{\PTJWSTPresetMedSigTwoFive}{902}
\newcommand{\PTJWSTPresetN}{20}
\newcommand{\PTJWSTPresetRawTotDelta}{0.0444}

\newcommand{\PTJWSTPresetSFourFive}{0.0837}
\newcommand{\PTJWSTPresetSFourFiveThreeDec}{0.084}
\newcommand{\PTJWSTPresetSGScatter}{0.0484}

\newcommand{\PTJWSTPresetSHiFourFive}{0.1028}
\newcommand{\PTJWSTPresetSHiOneFive}{0.0329}

\newcommand{\PTJWSTPresetSHiThreeFive}{0.0638}
\newcommand{\PTJWSTPresetSHiTwoFive}{0.0448}

\newcommand{\PTJWSTPresetSLoFourFive}{0.0539}
\newcommand{\PTJWSTPresetSLoOneFive}{0.0189}

\newcommand{\PTJWSTPresetSLoThreeFive}{0.0337}
\newcommand{\PTJWSTPresetSLoTwoFive}{0.0222}
\newcommand{\PTJWSTPresetSOneFive}{0.0277}

\newcommand{\PTJWSTPresetSThreeFive}{0.0519}
\newcommand{\PTJWSTPresetSTwoFive}{0.0358}
\newcommand{\PTJWSTPresetSTwoFiveThreeDec}{0.036}
\newcommand{\PTJWSTUnflaggedDelta}{0.0563}
\newcommand{\PTJWSTUnflaggedDeltaHi}{0.0715}
\newcommand{\PTJWSTUnflaggedDeltaLo}{0.0011}
\newcommand{\PTJWSTUnflaggedN}{8}

\newcommand{\PTMockFullJWSTGaussDeltaHi}{0.2081}
\newcommand{\PTMockFullJWSTGaussDeltaLo}{0.0256}

\newcommand{\PTMockFullJWSTWingDeltaHi}{0.1013}
\newcommand{\PTMockFullJWSTWingDeltaLo}{0.0185}

\newcommand{\PTMockPresetJWSTGaussDeltaHi}{0.2156}
\newcommand{\PTMockPresetJWSTGaussDeltaLo}{0.0215}
\newcommand{\PTMockPresetJWSTGaussDeltaMed}{0.0731}
\newcommand{\PTMockPresetJWSTGaussDeltaTrue}{0.0113}

\newcommand{\PTMockPresetJWSTGaussTrueFull}{0.0481}
\newcommand{\PTMockPresetJWSTGaussTrueTwoFive}{0.0364}
\newcommand{\PTMockPresetJWSTWingDeltaHi}{0.1069}
\newcommand{\PTMockPresetJWSTWingDeltaLo}{0.0143}
\newcommand{\PTMockPresetJWSTWingDeltaMed}{0.0416}
\newcommand{\PTMockPresetJWSTWingDeltaTrue}{0.0205}

\newcommand{\PTMockPresetSDSSGaussDeltaHi}{0.0776}
\newcommand{\PTMockPresetSDSSGaussDeltaLo}{0.0553}

\newcommand{\PTMockPresetSDSSGaussDeltaTrue}{0.0575}

\newcommand{\PTMockPresetSDSSGaussTrueFull}{0.1425}
\newcommand{\PTMockPresetSDSSGaussTrueTwoFive}{0.0619}
\newcommand{\PTMockPresetSDSSWingDeltaHi}{0.0409}
\newcommand{\PTMockPresetSDSSWingDeltaLo}{0.0293}

\newcommand{\PTMockPresetSDSSWingDeltaTrue}{0.0334}

\newcommand{\PTSDSSFullDeltaHi}{0.0250}
\newcommand{\PTSDSSFullDeltaLo}{0.0181}
\newcommand{\PTSDSSFullDeltaSigned}{+0.0217}

\newcommand{\PTSDSSFullNFourFive}{179}
\newcommand{\PTSDSSFullNOneFive}{181}
\newcommand{\PTSDSSFullNThreeFive}{180}
\newcommand{\PTSDSSFullNTwoFive}{181}
\newcommand{\PTSDSSFullSFourFive}{0.0742}
\newcommand{\PTSDSSFullSHiFourFive}{0.0793}
\newcommand{\PTSDSSFullSHiOneFive}{0.0364}
\newcommand{\PTSDSSFullSHiThreeFive}{0.0719}
\newcommand{\PTSDSSFullSHiTwoFive}{0.0565}
\newcommand{\PTSDSSFullSLoFourFive}{0.0679}
\newcommand{\PTSDSSFullSLoOneFive}{0.0297}
\newcommand{\PTSDSSFullSLoThreeFive}{0.0620}
\newcommand{\PTSDSSFullSLoTwoFive}{0.0482}
\newcommand{\PTSDSSFullSOneFive}{0.0332}
\newcommand{\PTSDSSFullSThreeFive}{0.0673}
\newcommand{\PTSDSSFullSTwoFive}{0.0527}

\newcommand{\PTSDSSPresetDeltaHi}{0.0231}
\newcommand{\PTSDSSPresetDeltaLo}{0.0167}
\newcommand{\PTSDSSPresetDeltaSigned}{+0.0200}

\newcommand{\PTSDSSPresetDeltaThreeDec}{0.020}

\newcommand{\PTSDSSPresetNFourFive}{167}
\newcommand{\PTSDSSPresetNOneFive}{169}

\newcommand{\PTSDSSPresetNThreeFive}{168}
\newcommand{\PTSDSSPresetNTwoFive}{169}

\newcommand{\PTSDSSPresetSFourFive}{0.0718}

\newcommand{\PTSDSSPresetSHiFourFive}{0.0769}
\newcommand{\PTSDSSPresetSHiOneFive}{0.0362}

\newcommand{\PTSDSSPresetSHiThreeFive}{0.0710}
\newcommand{\PTSDSSPresetSHiTwoFive}{0.0562}

\newcommand{\PTSDSSPresetSLoFourFive}{0.0655}
\newcommand{\PTSDSSPresetSLoOneFive}{0.0293}

\newcommand{\PTSDSSPresetSLoThreeFive}{0.0607}
\newcommand{\PTSDSSPresetSLoTwoFive}{0.0473}
\newcommand{\PTSDSSPresetSOneFive}{0.0330}

\newcommand{\PTSDSSPresetSThreeFive}{0.0663}
\newcommand{\PTSDSSPresetSTwoFive}{0.0521}

\newcommand{\SDSSBridgeN}{59}
\newcommand{\SDSSCleanLHaMax}{43.57}
\newcommand{\SDSSCleanLHaMed}{42.86}
\newcommand{\SDSSCleanLHaMin}{42.15}
\newcommand{\SDSSCleanMedFWHM}{3396}
\newcommand{\SDSSCleanN}{185}
\newcommand{\SDSSCleanSGScatterUniform}{0.1465}

\newcommand{\SDSSCleanZMax}{0.38}
\newcommand{\SDSSCleanZMed}{0.29}
\newcommand{\SDSSCleanZMin}{0.12}

\newcommand{\SDSSSpectra}{204}

\newcommand{\Ha}{\ensuremath{\mathrm{H}\alpha}}
\newcommand{\Hb}{\ensuremath{\mathrm{H}\beta}}
\newcommand{\OIII}{\ensuremath{[\mathrm{O\,\textsc{iii}}]}}
\newcommand{\NII}{\ensuremath{[\mathrm{N\,\textsc{ii}}]}}
\newcommand{\SII}{\ensuremath{[\mathrm{S\,\textsc{ii}}]}}
\newcommand{\OI}{\ensuremath{[\mathrm{O\,\textsc{i}}]}}
\newcommand{\HeI}{\ensuremath{\mathrm{He\,\textsc{i}}}}
\newcommand{\kms}{\ensuremath{\mathrm{km\,s^{-1}}}}
\newcommand{\LHa}{\ensuremath{L_{\mathrm{H}\alpha}}}
\newcommand{\FWHM}{\ensuremath{\mathrm{FWHM}}}
\newcommand{\FWHMsg}{\ensuremath{\mathrm{FWHM}_{\mathrm{sg}}}}
\newcommand{\fvir}{\ensuremath{f_{\mathrm{vir}}}}
\newcommand{\MBH}{\ensuremath{M_{\mathrm{BH}}}}
\newcommand{\sigV}{\ensuremath{\sigma(V)}}

\shorttitle{Velocity-Window Dependence of the LRD \Ha\ Relation}
\shortauthors{Wang}
\hypersetup{pdftitle={Velocity-Window Dependence of the Broad H-alpha Width-Luminosity Relation in Little Red Dots},pdfauthor={Haoyu Wang}}

\begin{document}

\title{Velocity-Window Dependence of the Broad \texorpdfstring{\Ha}{H-alpha} Width--Luminosity Relation in Little Red Dots}

\correspondingauthor{Haoyu Wang}
\email{pkuwanghaoyu@pku.edu.cn}

\author[0009-0005-9117-0842]{Haoyu Wang}
\affiliation{Independent Researcher}

\begin{abstract}
Broad Balmer emission lines probe black hole masses and nuclear kinematics in high-redshift active galactic nuclei (AGNs). In JWST ``little red dots'' (LRDs), integrated broad-\Ha\ widths span a remarkably narrow range at fixed line luminosity, raising the question of whether this tightness persists once outer-wing emission is included. Holding each object and its broad-\Ha\ luminosity fixed, we measure the finite-window line dispersion \sigV, the flux-weighted second-moment width within $\pm V$, from $V = 1500$ to $4500~\kms$ in JWST/NIRSpec spectra and Sloan Digital Sky Survey (SDSS) Type-1 AGNs. Across $\PTJWSTPresetN$ high signal-to-noise LRDs, the observed scatter in $\log_{10}\sigV$ at fixed \LHa\ rises from $\PTJWSTPresetSTwoFiveThreeDec$ to $\PTJWSTPresetSFourFiveThreeDec$~dex between $V = 2500$ and $4500~\kms$, a paired increase of $\PTJWSTPresetDeltaThreeDec$~dex with a 95\% bootstrap interval of $\PTJWSTPresetDeltaLoThreeDec$--$\PTJWSTPresetDeltaHiThreeDec$~dex, compared with $\PTSDSSPresetDeltaThreeDec$~dex across $\PTSDSSPresetNFourFive$ broader SDSS AGNs. Injecting synthetic profiles into the empirical noise of the same $\PTJWSTPresetN$ spectra shows that single-Gaussian and core-plus-wing models both accommodate this rise, leaving intrinsic wing diversity, truncation, and wing noise unseparated. In the clearest dual-line spectrum, RUBIES-EGS-55604, \Hb\ strengthens relative to \Ha\ from line center to wing under two spectral decompositions of the same grating spectrum. Alongside integrated widths in JWST, SDSS, and $\EuclidMainN$ Euclid DR1 AGNs, these results show that the observed tightness of the \Ha\ scaling depends on the velocities included in \sigV, requiring mass and kinematic models to match both transition and velocity interval.
\end{abstract}


\section{Introduction}
\label{sec:intro}

JWST spectroscopy has revealed an abundant population of compact rest-optical sources at $z \gtrsim 3$, termed ``little red dots'' (LRDs), with prominent broad Balmer emission lines \citep{Matthee2024,Greene2024} and characteristic V-shaped continua inflecting near the Balmer limit \citep{Setton2025Vshape}. In several bright examples with simultaneous \Ha\ and \Hb\ grating coverage, \citet[][Section~V.3.1 and Figure~9]{Chen2026} further showed that the $F(\Ha)/F(\Hb)$ flux ratio decreases substantially from line center toward the wings, indicating that emission or radiative-transfer conditions vary across an individual broad-line profile.

Whether LRD broad lines trace virialized motion around massive black holes or radiative transfer in dense gas envelopes remains actively debated \citep[e.g.,][]{Rusakov2026,Madau2026b,InayoshiHo}. In classical active galactic nuclei (AGNs), single-epoch black hole mass estimates follow $\MBH = \fvir\, R\, \Delta V^2 / G$ \citep{Peterson2004,GreeneHo2005,Bentz2013}, where the virial factor \fvir\ depends on whether $\Delta V$ is characterized by the full width at half maximum (\FWHM) or the flux-weighted line dispersion \citep{Collin2006}. Across AGN populations, width scatter at fixed luminosity constrains broad-line geometry and the dynamical information carried by single-epoch widths \citep{Fine2008,Croom2011,Assef2012}, while optical-depth effects and scattering can reshape emergent line wings relative to bulk gas velocities \citep{OsterbrockFerland2006,Laor2006}.

Whereas local quasars typically show $\gtrsim 0.12$~dex of width scatter at fixed luminosity, \citet{Wang2026} found that JWST LRDs follow a tight broad-\Ha\ width--luminosity relation with a measurement-error-corrected intrinsic scatter of $0.054$~dex at fixed \LHa, less than half that of refitted SDSS quasars and independently seen in \citet{BaccusXu2026}. \citet{Wang2026} also found similarly small scatter for single-Gaussian \FWHM\ (\FWHMsg), interpercentile width $W_{50}$, and Lorentzian or exponential fits, alongside larger scatter for two-Gaussian composite \FWHM. \FWHM\ is set by the profile near half maximum, whereas flux at large velocity offsets carries greater weight in the second moment; changes in the wings can therefore affect the two width measures differently, so switching between full-profile width estimators does not isolate individual velocity ranges. A direct test instead requires varying the velocity interval itself. Here we hold the objects, width definition, and broad-\Ha\ luminosities fixed, varying only the velocity range $[-V, +V]$ in the finite-window line dispersion \sigV\ to test whether LRD widths remain tightly clustered once outer wings are included.

Section~\ref{sec:res_moments} and Figure~\ref{fig:moments} measure how the observed \Ha\ scatter changes across $V = 1500$--$4500~\kms$ in JWST LRDs and SDSS Type-1 AGNs, calibrated by empirical-noise profile injections on the same targets. Section~\ref{sec:res_balmer} and Figure~\ref{fig:balmer} examine velocity-binned $F(\Hb)/F(\Ha)$ ratios under two decompositions of the same combined grating spectra in three high-weight dual-line LRDs and relate them to a sample-wide template fit. Section~\ref{sec:res_context} and Figure~\ref{fig:three_survey} place integrated LRD \Ha\ widths alongside SDSS and Euclid DR1 AGNs, and Section~\ref{sec:discussion} discusses implications for broad-line modeling.

\section{Observations and Samples}
\label{sec:samples}

Table~\ref{tab:samples} summarizes all spectroscopic samples and subsets analyzed here. Because the three surveys span disjoint regimes of redshift, spectral resolution, and continuum luminosity, each serves a distinct empirical role rather than forming a fixed-luminosity evolutionary sequence.

\begin{deluxetable*}{lcccp{3.75cm}}
\tabletypesize{\footnotesize}
\tabcolsep=3.8pt
\tablewidth{0pt}
\tablecaption{Samples and Measurements\label{tab:samples}}
\tablehead{
\colhead{Survey / Subset} &
\colhead{$N$} &
\colhead{Redshift} &
\colhead{Measured Quantities} &
\colhead{Selection and Analytical Role}
}
\startdata
\cutinhead{JWST/NIRSpec Little Red Dots (Archival Grating Parent: $N = \JWSTGratingParentN$)}
Continuum-clean baseline &
  $\JWSTCleanN$\tablenotemark{a} &
  $\JWSTCleanZMin$--$\JWSTCleanZMax$ &
  \Ha\ \FWHMsg &
  Continuum-verified integrated \Ha\ width--luminosity baseline \\
Primary window sample &
  $\PTJWSTPresetN$\tablenotemark{b} &
  $\JWSTPresetZMin$--$\JWSTPresetZMax$ &
  \Ha\ \sigV, \FWHMsg &
  Broad \Ha\ $\mathrm{S/N} \ge 15$ with both default continuum sidebands \\
Supplementary window sample &
  $\PTJWSTFullN$\tablenotemark{b} &
  $\JWSTHighSNRZMin$--$\JWSTHighSNRZMax$ &
  \Ha\ \sigV, \FWHMsg &
  Adds $\JWSTEdgeFallbackN$ detector-edge sources via near-line red continuum sideband \\
Same-grating Balmer sample &
  $\POneTmplAllThirtyThreeN$ (top $\POneTmplTopThreeN$)\tablenotemark{c} &
  $\JWSTTwoLineZMin$--$\JWSTTwoLineZMax$ &
  \Hb/\Ha\ $R(v)$, template &
  Simultaneous \Ha\ and \Hb\ on same combined grating spectrum \\
\cutinhead{Comparison and Population-Context Surveys}
SDSS DR7 Type-1 AGNs &
  $\SDSSCleanN$ ($\PTSDSSPresetNFourFive$--$\PTSDSSPresetNOneFive$)\tablenotemark{d} &
  $\SDSSCleanZMin$--$\SDSSCleanZMax$ &
  \Ha\ \sigV, \FWHMsg &
  Same-window empirical reference in local broad-line AGNs \\
Euclid DR1 primary sample &
  $\EuclidMainN$\tablenotemark{e} &
  $\EuclidMainZMin$--$\EuclidMainZMax$ &
  \Ha\ \FWHMsg, $\sigma_w$ &
  Independently selected intermediate-$z$ integrated-width reference \\
\enddata
\tablenotetext{a}{Subset of the $\JWSTGratingParentN$ archival NIRSpec grating LRDs passing local continuum verification around \Ha.}
\tablenotetext{b}{Among the $\PTJWSTFullN$ continuum-clean sources with broad-\Ha\ $\mathrm{S/N} \ge 15$, $\PTJWSTPresetN$ cover both default sidebands (primary) and $\JWSTEdgeFallbackN$ are detector-edge objects (supplementary).}
\tablenotetext{c}{Selected directly from the $\JWSTGratingParentN$ grating sources, comprising $\JWSTTwoLineInCleanN$ of the $\JWSTCleanN$ continuum-clean objects and $\JWSTTwoLineOutCleanN$ objects outside that subset; velocity-binned ratios are analyzed in the $\POneTmplTopThreeN$ highest-weight objects.}
\tablenotetext{d}{From $\SDSSSpectra$ spectra cross-matched to \citet{Shen2011}, $\SDSSCleanN$ pass broad-\Ha\ refit quality control and $\PTSDSSPresetNFourFive$--$\PTSDSSPresetNOneFive$ yield valid moments across $V = 1500$--$4500~\kms$ under default sidebands, rising to $\PTSDSSFullNFourFive$--$\PTSDSSFullNOneFive$ with near-line sidebands; see Table~\ref{tab:moments}.}
\tablenotetext{e}{Selected from $\EuclidTargets$ Euclid DR1 candidates at $19.0 < H_{\mathrm{aper}} < \EuclidHLimit$ with usable MER photometry prior to inspecting broad-line widths.}
\end{deluxetable*}

\emph{JWST/NIRSpec LRDs.}---We start from an archival parent compilation of $\JWSTGratingParentN$ deduplicated LRDs observed with JWST/NIRSpec medium- or high-resolution gratings at $R \sim 1000$--$2700$ \citep{Jakobsen2022,deGraaffLSF2024}. Reduced 1D spectra were retrieved from the DAWN JWST Archive,\footnote{\url{https://dawn-cph.github.io/dja/}} extracted with \texttt{msaexp} \citep{msaexp} and including public observations from the RUBIES survey \citep{DJAdeGraaff2024}. Our target list combines published spectroscopic LRD compilations, primarily \citet{BaccusXu2026} and Table~1 of \citet{Yanagisawa2026} alongside \citet{Kocevski2025}, \citet{Taylor2025}, and \citet{Hviding2025}, retaining any source classified as an LRD in at least one contributing catalog. Cross-matching against the full JWST spectroscopic sample of \citet{Wang2026} yields $\JWSTGratingParentWangOverlapN$ shared sources among the $\JWSTGratingParentN$ parent targets, corresponding to $\JWSTCleanWangOverlapN$ of the $\JWSTCleanN$ continuum-clean sources, $\JWSTHighSNRWangOverlapN$ of the $\JWSTHighSNRN$ high-S/N sources, $\JWSTPresetWangOverlapN$ of the $\PTJWSTPresetN$ primary window sources, and $\JWSTTwoLineWangOverlapN$ of the $\POneTmplAllThirtyThreeN$ dual-line sources; these overlap counts include CEERS-1244, which served as a comparison AGN in \citet{Wang2026} but is classified as an LRD in Table~1 of \citet{Yue2024}. The $\JWSTGratingParentN$ grating spectra enter two analyses:
\begin{enumerate}
\item \emph{\Ha\ velocity-window sample.}---Of the $\JWSTGratingParentN$ grating sources, $\JWSTCleanN$ at $z = \JWSTCleanZMin$--$\JWSTCleanZMax$ (median $\JWSTCleanZMed$, including $\JWSTCleanBXProvN$ in \citealt{BaccusXu2026} and $\JWSTCleanAtlasProvN$ in \citealt{Yanagisawa2026}) pass local continuum verification for the integrated-width baseline. Single-Gaussian fits define the uniform comparison width \FWHMsg\ with median $\JWSTCleanMedFWHM~\kms$, whereas broad-\Ha\ luminosities, spanning $\log_{10}(\LHa/\mathrm{erg\,s^{-1}}) = \JWSTCleanLHaMin$--$\JWSTCleanLHaMax$ with median $\JWSTCleanLHaMed$, and broad-line S/N use the total broad flux selected by the Bayesian information criterion (BIC); a symmetric exponential or summed two-Gaussian profile is adopted only when it yields the lowest BIC by $\Delta\mathrm{BIC}>6$ relative to the single Gaussian, and each object's luminosity is held fixed across all velocity windows. Within these $\JWSTCleanN$ objects, $\JWSTHighSNRN$ satisfy $\mathrm{S/N} \ge 15$ across $z = \JWSTHighSNRZMin$--$\JWSTHighSNRZMax$ with median $\FWHMsg = \JWSTHighSNRMedFWHM~\kms$. Both default continuum sidebands are covered in $\PTJWSTPresetN$ of these $\JWSTHighSNRN$ spectra, forming our \emph{primary} window sample, while adding $\JWSTEdgeFallbackN$ detector-edge targets, RUBIES-EGS-50052 and JADES-GS-38562, via a near-line red continuum sideband defines the $\PTJWSTFullN$-source \emph{supplementary} sample detailed in Appendix~\ref{app:edge}.
\item \emph{Dual-line Balmer sample.}---Separately, $\POneTmplAllThirtyThreeN$ of the $\JWSTGratingParentN$ grating sources at $z = \JWSTTwoLineZMin$--$\JWSTTwoLineZMax$ cover \Ha\ and \Hb\ simultaneously on the same combined grating spectrum, comprising $\JWSTTwoLineInCleanN$ of the $\JWSTCleanN$ continuum-clean sources and $\JWSTTwoLineOutCleanN$ additional targets, RUBIES-UDS-149298, Skyfire-1090, and RUBIES-EGS-42803. Velocity-binned Balmer ratios are examined in the $\POneTmplTopThreeN$ objects carrying the highest inverse-variance weight in the $\POneTmplAllThirtyThreeN$-source two-line template fit: RUBIES-EGS-55604, RUBIES-EGS-42046, and JADES-GN-68797.
\end{enumerate}

\emph{SDSS DR7 Type-1 AGNs.}---As a low-redshift empirical reference measured with the identical finite-window pixel integration, we use $\SDSSCleanN$ quality-controlled broad-line AGNs at $z = \SDSSCleanZMin$--$\SDSSCleanZMax$ (median $\SDSSCleanZMed$, median $\FWHMsg = \SDSSCleanMedFWHM~\kms$) refitted from $\SDSSSpectra$ SDSS DR7 spectra cross-matched to \citet{Shen2011}, requiring a converged broad-\Ha\ refit with broad-line $\mathrm{S/N} \ge 5$, $\Delta\mathrm{BIC} \ge 6$ over a narrow-only model, and fitted core width within $505$--$9400~\kms$. We convert the cataloged broad-\Ha\ luminosities of \citet{Shen2011} to our adopted cosmology, spanning $\log_{10}(\LHa/\mathrm{erg\,s^{-1}}) = \SDSSCleanLHaMin$--$\SDSSCleanLHaMax$ with median $\SDSSCleanLHaMed$. Default continuum sidebands yield $\PTSDSSPresetNFourFive$--$\PTSDSSPresetNOneFive$ valid spectra across $V = 1500$--$4500~\kms$, increasing to $\PTSDSSFullNFourFive$--$\PTSDSSFullNOneFive$ when near-line sidebands are allowed for edge spectra (Table~\ref{tab:moments}).

\emph{Euclid DR1 Primary Sample and Cosmology.}---To establish an independently selected intermediate-redshift reference, we use $\EuclidMainN$ broad-\Ha\ AGNs at $z = \EuclidMainZMin$--$\EuclidMainZMax$ (median $\EuclidMainZMed$, median $\FWHMsg = \EuclidMainMedFWHM~\kms$, and $\log_{10}(\LHa/\mathrm{erg\,s^{-1}}) = \EuclidMainLHaMin$--$\EuclidMainLHaMax$) measured from Euclid DR1 NISP slitless spectroscopy \citep{EuclidAussel2026,EuclidLeBrun2026,EuclidFumana2026}. We selected the parent list of $\EuclidTargets$ candidates at $19.0 < H_{\mathrm{aper}} < \EuclidHLimit$ with usable MER photometry prior to inspecting the broad-line widths---distinct from photometric LRD selections in Euclid Quick Data Release~1 \citep{EuclidBisigello2025}---retaining sources with $\ge 3$ dithers, $\ge 85\%$ valid pixels and $\ge 70\%$ paired grism-orientation coverage within $\pm 400$~\AA\ of \Ha, a converged broad-\Ha\ fit with deconvolved $\FWHM \ge 1000~\kms$ at $\mathrm{S/N} \ge 10$, and positive line-subtracted MER $J_E/H_E$ continua. Because NISP slitless resolution and per-pixel S/N do not support native-pixel wing moments out to $\pm 4500~\kms$, Euclid enters solely through integrated-width statistics. Across the shared luminosity band $\log_{10}(\LHa/\mathrm{erg\,s^{-1}}) = 43.0$--$43.7$, SDSS and Euclid contribute $\SDSSBridgeN$ and $\EuclidBridgeN$ objects whereas JWST contributes only $\JWSTBridgeN$, and the brightest JWST overlap continuum at $\log_{10}(\lambda L_\lambda/\mathrm{erg\,s^{-1}}) = \JWSTBridgeCMax$ lies $\BridgeJWSTEuclidContinuumGap$~dex below the faintest Euclid overlap source at $\EuclidBridgeCMin$. Throughout this paper, luminosity distances use the \texttt{Planck18} flat $\Lambda$CDM cosmology in Astropy with $H_0 = 67.66~\kms\,\mathrm{Mpc}^{-1}$ and $\Omega_{\mathrm{m},0} = 0.30966$ \citep{Planck2020}.

\section{Measurements and Analysis}
\label{sec:methods}

\subsection{Finite-Window Line Dispersion}
\label{sec:methods_moments}

For each JWST and SDSS spectrum in the \Ha\ moment scan, rest-frame velocity $v$ is referenced to the fitted broad-\Ha\ center. A linear inverse-variance continuum is fitted across default two-sided sidebands at $[-11000, -6500]$ and $[+8700, +11000]~\kms$, masking pixels within $450~\kms$ of rest-frame $\OI\,\lambda 6365$ and avoiding $\HeI\,\lambda 6678$ and \SII. Within $|v| \le 4500~\kms$, a two-Gaussian broad model plus narrow \Ha\ and $\NII\,\lambda\lambda 6548,6583$ are fitted simultaneously solely to subtract the narrow lines. We retain the signed narrow-subtracted residual flux $f(v)$ on the native pixel grid, preserving Balmer absorption troughs and each instrument's native line-spread function (LSF), and center the moment on the flux centroid $\bar{v}_V$ within each symmetric window $[-V, +V]$:
\begin{equation}
\sigma^2(V) = \frac{\int_{-V}^{V} (v - \bar{v}_V)^2 f(v)\,dv}{\int_{-V}^{V} f(v)\,dv},
\quad
\bar{v}_V = \frac{\int_{-V}^{V} v\, f(v)\,dv}{\int_{-V}^{V} f(v)\,dv}.
\label{eq:sigma_v}
\end{equation}
We refer to $\sigV \equiv \sqrt{\sigma^2(V)}$ throughout as the \emph{finite-window line dispersion}. Exact positive-overlap pixel quadrature formulas are given in Appendix~\ref{app:exact_int}; to avoid bridging wavelength gaps, we cap each pixel half-width at $0.55$ times the median pixel spacing and require $\ge 98\%$ valid velocity coverage, positive integrated flux, and $\sigma^2(V) > 0$.

Two properties govern \sigV\ across $V \in \{1500, 2500, 3500, 4500\}~\kms$. First, for a nearly flat flux density inside a narrow window, Equation~(\ref{eq:sigma_v}) approaches $\sigV = V/\sqrt{3}$ as derived in Appendix~\ref{app:exact_int}, compressing differences between broader and narrower profiles; in addition, a fixed velocity cut $V$ reaches further into the wings of narrower LRD profiles, which have median $\FWHMsg = \JWSTCleanMedFWHM~\kms$, than of broader SDSS profiles with median $\SDSSCleanMedFWHM~\kms$. Second, the $(v - \bar{v}_V)^2$ factor amplifies weak-wing noise and continuum-placement errors at large $V$, a sensitivity familiar from repeated single-epoch AGN spectroscopy \citep{Denney2009}. In empirical four-pixel block-noise recovery tests defined prior to measuring the observed moments (Appendix~\ref{app:injections}), high-S/N JWST spectra satisfy all pooled qualification thresholds through $V = 4500~\kms$, achieving 68th-percentile recovery errors of $\PTCalJWSTGaussFourFiveErrSixEight$ and $\PTCalJWSTWingFourFiveErrSixEight$~dex for Gaussian and core-plus-wing inputs, respectively, before losing joint qualification at $V = 5500$ and $6000~\kms$.

\subsection{Velocity-Binned Balmer Flux Ratios}
\label{sec:methods_balmer}

We measure resolved Balmer profiles in RUBIES-EGS-55604 at $z_{\OIII} = \POneSrcFiveFiveZOIII$ (G395M, $\POneSrcFiveFiveNCombine$ combined exposures), RUBIES-EGS-42046 at $z_{\OIII} = \POneSrcFourTwoZOIII$ (G395M, $\POneSrcFourTwoNCombine$ combined exposures), and JADES-GN-68797 at $z_{\OIII} = \POneSrcSixEightZOIII$ (G395H, $\POneSrcSixEightNCombine$ combined exposures). Velocities are referenced to local systemic \OIII, rather than the broad-\Ha\ centroid used in Section~\ref{sec:methods_moments} and in the $\POneTmplAllThirtyThreeN$-source two-line template, and native-pixel overlaps are integrated across five symmetric $|v|$ intervals: $0$--$300$, $300$--$1000$, $1000$--$1500$, $1500$--$2500$, and $2500$--$3500~\kms$. On each combined grating spectrum, \Ha\ is convolved with a velocity-space Gaussian kernel matching the local \Hb\ LSF before evaluating $R \equiv F(\Hb)/F(\Ha)$ after joint continuum and narrow-line subtraction, denoted decomposition~A, and after subtracting only an independent local linear continuum around each line, denoted decomposition~B. Both decompositions retain Balmer absorption without infill; equivalent Balmer decrements $D = 1/R$ are quoted only where integrated $\Hb \ge 2\sigma$.

\subsection{Residual Scatter and Resampling}
\label{sec:methods_regression}

At each window half-width $V$, we report $s(V)$ in dex as the sample standard deviation (with $N - 1$ degrees of freedom) of the residuals about a separate ordinary least-squares (OLS) linear regression of $\log_{10}\sigV$, or $\log_{10}\FWHMsg$ for integrated widths, on $\log_{10}\LHa$. Single-window 95\% confidence intervals take the equal-tailed 2.5th--97.5th percentiles across 5000 bootstrap resamples of valid sources with replacement, refitting the relation on each draw. For the window-to-window change $\Delta s_{2500\to 4500} \equiv s(4500) - s(2500)$, we resample objects measured in both windows as pairs across 5000 draws and refit both relations on every draw, whereas cross-survey scatter differences resample each survey independently across 5000 draws. Throughout, $s(V)$ and $\Delta s_{2500\to 4500}$ measure \emph{observed} residual scatter rather than measurement-error-corrected intrinsic scatter, and \LHa-conditioned scatter is kept distinct from Euclid's continuum-conditioned, selection-corrected intrinsic scatter $\sigma_w$.

\section{Results}
\label{sec:results}

\subsection{Velocity-Window Response of the Observed \texorpdfstring{\Ha}{H-alpha} Scatter}
\label{sec:res_moments}

\begin{figure*}[t]
\centering
\includegraphics[width=\textwidth]{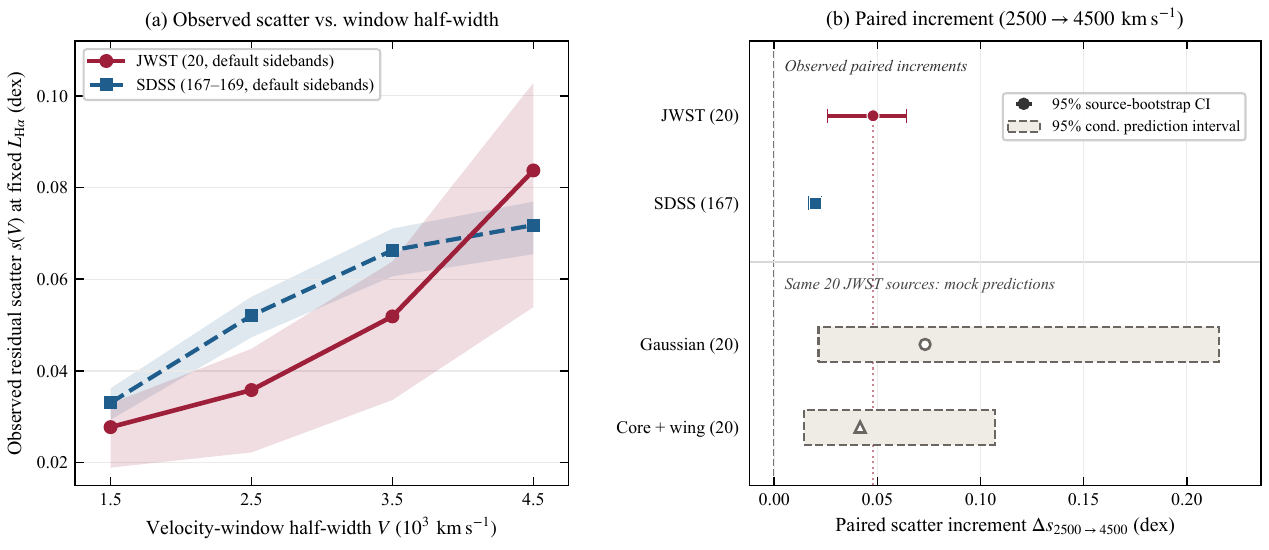}
\caption{Observed scatter in broad-\Ha\ finite-window line dispersion \sigV\ at fixed \LHa\ under unified default continuum sidebands. \textbf{(a)} Per-window OLS residual scatter $s(V)$ across the validated range $V = 1500$--$4500~\kms$ for the $\PTJWSTPresetN$ primary JWST LRDs (crimson circles and solid curve) and $\PTSDSSPresetNFourFive$--$\PTSDSSPresetNOneFive$ SDSS Type-1 AGNs (steel-blue squares and dashed curve); shaded bands denote per-window 95\% source-bootstrap confidence intervals, with extended-window diagnostics in Appendix~\ref{app:calibration}. \textbf{(b)} Paired scatter increment $\Delta s_{2500\to 4500}$ from $V = 2500$ to $4500~\kms$. Upper rows show observed paired increments with 95\% source-bootstrap confidence intervals (filled symbols with capped solid bars) for the $\PTJWSTPresetN$ JWST sources and the $\PTSDSSPresetNFourFive$ SDSS sources valid in both windows. Lower rows show matching 95\% conditional prediction intervals (capless shaded dashed bands with open median markers) from synthetic profile injections into the empirical noise of the same $\PTJWSTPresetN$ JWST spectra for single-Gaussian and core-plus-wing models; the vertical dotted crimson line marks the observed $\PTJWSTPresetN$-source JWST increment.}
\label{fig:moments}
\end{figure*}

\begin{deluxetable*}{lccc}
\tabletypesize{\footnotesize}
\tablewidth{0pt}
\tablecaption{Observed Residual Scatters $s(V)$ and Paired Increments Across Validated Velocity Windows\label{tab:moments}}
\tablehead{
\colhead{Window Half-Width / Quantity} &
\colhead{JWST (dex)} &
\colhead{SDSS (dex)} &
\colhead{SDSS $N$}
}
\startdata
\cutinhead{Primary Comparison: Default Continuum Sidebands (JWST $N = \PTJWSTPresetN$)}
$V = 1500~\kms$ &
  $\PTJWSTPresetSOneFive$ $[\PTJWSTPresetSLoOneFive, \PTJWSTPresetSHiOneFive]$ &
  $\PTSDSSPresetSOneFive$ $[\PTSDSSPresetSLoOneFive, \PTSDSSPresetSHiOneFive]$ &
  $\PTSDSSPresetNOneFive$ \\
$V = 2500~\kms$ &
  $\PTJWSTPresetSTwoFive$ $[\PTJWSTPresetSLoTwoFive, \PTJWSTPresetSHiTwoFive]$ &
  $\PTSDSSPresetSTwoFive$ $[\PTSDSSPresetSLoTwoFive, \PTSDSSPresetSHiTwoFive]$ &
  $\PTSDSSPresetNTwoFive$ \\
$V = 3500~\kms$ &
  $\PTJWSTPresetSThreeFive$ $[\PTJWSTPresetSLoThreeFive, \PTJWSTPresetSHiThreeFive]$ &
  $\PTSDSSPresetSThreeFive$ $[\PTSDSSPresetSLoThreeFive, \PTSDSSPresetSHiThreeFive]$ &
  $\PTSDSSPresetNThreeFive$ \\
$V = 4500~\kms$ &
  $\PTJWSTPresetSFourFive$ $[\PTJWSTPresetSLoFourFive, \PTJWSTPresetSHiFourFive]$ &
  $\PTSDSSPresetSFourFive$ $[\PTSDSSPresetSLoFourFive, \PTSDSSPresetSHiFourFive]$ &
  $\PTSDSSPresetNFourFive$ \\
Paired $\Delta s_{2500\to 4500}$ &
  $\PTJWSTPresetDeltaSigned$ $[\PTJWSTPresetDeltaLo, \PTJWSTPresetDeltaHi]$ &
  $\PTSDSSPresetDeltaSigned$ $[\PTSDSSPresetDeltaLo, \PTSDSSPresetDeltaHi]$ &
  $\PTSDSSPresetNFourFive$ \\
\cutinhead{Supplementary Check: Near-Line Sidebands Allowed for Edge Spectra (JWST $N = \PTJWSTFullN$)}
$V = 1500~\kms$ &
  $\PTJWSTFullSOneFive$ $[\PTJWSTFullSLoOneFive, \PTJWSTFullSHiOneFive]$ &
  $\PTSDSSFullSOneFive$ $[\PTSDSSFullSLoOneFive, \PTSDSSFullSHiOneFive]$ &
  $\PTSDSSFullNOneFive$ \\
$V = 2500~\kms$ &
  $\PTJWSTFullSTwoFive$ $[\PTJWSTFullSLoTwoFive, \PTJWSTFullSHiTwoFive]$ &
  $\PTSDSSFullSTwoFive$ $[\PTSDSSFullSLoTwoFive, \PTSDSSFullSHiTwoFive]$ &
  $\PTSDSSFullNTwoFive$ \\
$V = 3500~\kms$ &
  $\PTJWSTFullSThreeFive$ $[\PTJWSTFullSLoThreeFive, \PTJWSTFullSHiThreeFive]$ &
  $\PTSDSSFullSThreeFive$ $[\PTSDSSFullSLoThreeFive, \PTSDSSFullSHiThreeFive]$ &
  $\PTSDSSFullNThreeFive$ \\
$V = 4500~\kms$ &
  $\PTJWSTFullSFourFive$ $[\PTJWSTFullSLoFourFive, \PTJWSTFullSHiFourFive]$ &
  $\PTSDSSFullSFourFive$ $[\PTSDSSFullSLoFourFive, \PTSDSSFullSHiFourFive]$ &
  $\PTSDSSFullNFourFive$ \\
Paired $\Delta s_{2500\to 4500}$ &
  $\PTJWSTFullDeltaSigned$ $[\PTJWSTFullDeltaLo, \PTJWSTFullDeltaHi]$ &
  $\PTSDSSFullDeltaSigned$ $[\PTSDSSFullDeltaLo, \PTSDSSFullDeltaHi]$ &
  $\PTSDSSFullNFourFive$ \\
\enddata
\tablecomments{Entries give the sample standard deviation of $\log_{10}\sigV$ residuals at fixed $\log_{10}\LHa$, uncorrected for measurement error, with 95\% source-bootstrap confidence intervals in brackets. Paired increments resample sources valid at both $V = 2500$ and $4500~\kms$, so the SDSS paired $\Delta s$ differs slightly from subtracting single-window scatters with unequal $N$. Matching 95\% conditional prediction intervals for $\Delta s_{2500\to 4500}$ are $[\PTMockPresetJWSTGaussDeltaLo, \PTMockPresetJWSTGaussDeltaHi]$~dex (Gaussian, median $\PTMockPresetJWSTGaussDeltaMed$) and $[\PTMockPresetJWSTWingDeltaLo, \PTMockPresetJWSTWingDeltaHi]$~dex (core + wing, median $\PTMockPresetJWSTWingDeltaMed$) for the $\PTJWSTPresetN$ default-sideband JWST sources, and $[\PTMockFullJWSTGaussDeltaLo, \PTMockFullJWSTGaussDeltaHi]$ and $[\PTMockFullJWSTWingDeltaLo, \PTMockFullJWSTWingDeltaHi]$~dex for the $\PTJWSTFullN$ JWST sources when near-line sidebands are included. In the $\PTJWSTFullN$-source comparison, per-window cross-survey differences $s_{\mathrm{JWST}}(V) - s_{\mathrm{SDSS}}(V)$ are $\PTDiffTwoFive$~$[\PTDiffLoTwoFive, \PTDiffHiTwoFive]$, $\PTDiffThreeFive$~$[\PTDiffLoThreeFive, \PTDiffHiThreeFive]$, and $\PTDiffFourFive$~$[\PTDiffLoFourFive, \PTDiffHiFourFive]$~dex at $V = 2500$, $3500$, and $4500~\kms$.}
\end{deluxetable*}

In the $\PTJWSTPresetN$ primary JWST LRDs, widening the \Ha\ velocity half-window from $2500$ to $4500~\kms$ more than doubles the observed scatter in finite-window line dispersion at fixed \LHa, increasing $s(V)$ from $\PTJWSTPresetSTwoFive$~dex to $\PTJWSTPresetSFourFive$~dex as the median $\sigV$ grows from $\PTJWSTPresetMedSigTwoFive$ to $\PTJWSTPresetMedSigFourFive~\kms$ (Figure~\ref{fig:moments}a; Table~\ref{tab:moments}). Resampling the $\PTJWSTPresetN$ objects as pairs yields $\Delta s_{2500\to 4500} = \PTJWSTPresetDeltaSigned$~dex with a 95\% source-bootstrap confidence interval of $[\PTJWSTPresetDeltaLo, \PTJWSTPresetDeltaHi]$~dex (Figure~\ref{fig:moments}b). A comparable increase appears when moments are integrated directly on the total continuum-subtracted \Ha\ profile without narrow-line removal, giving $\Delta s_{\mathrm{tot}} = +\PTJWSTPresetRawTotDelta$~dex across the same $\PTJWSTPresetN$ sources, and within the $\PTJWSTUnflaggedN$ high-S/N LRDs without flagged Balmer absorption, which show $+\PTJWSTUnflaggedDelta$~dex with a 95\% interval of $[\PTJWSTUnflaggedDeltaLo, \PTJWSTUnflaggedDeltaHi]$~dex.

Under the same default continuum sidebands and retaining each survey's native LSF, the SDSS Type-1 AGN reference rises more gradually across the identical velocity windows, giving a paired increment of $\Delta s_{2500\to 4500} = \PTSDSSPresetDeltaSigned$~dex with a 95\% confidence interval of $[\PTSDSSPresetDeltaLo, \PTSDSSPresetDeltaHi]$~dex across the $\PTSDSSPresetNFourFive$ spectra valid in both windows. Allowing near-line continuum sidebands for detector-edge spectra (lower block of Table~\ref{tab:moments}; Appendix~\ref{app:edge}) yields consistent paired increments of $\PTJWSTFullDeltaSigned$~dex ($[\PTJWSTFullDeltaLo, \PTJWSTFullDeltaHi]$~dex) for the $\PTJWSTFullN$ JWST sources and $\PTSDSSFullDeltaSigned$~dex ($[\PTSDSSFullDeltaLo, \PTSDSSFullDeltaHi]$~dex) for $\PTSDSSFullNFourFive$ paired SDSS sources. In that $\PTJWSTFullN$-source comparison, the JWST scatter remains smaller than the SDSS scatter at $V = 2500$ and $3500~\kms$, whereas at $4500~\kms$ the cross-survey difference interval spans zero at $\PTDiffFourFive$~dex ($[\PTDiffLoFourFive, \PTDiffHiFourFive]$~dex; Table~\ref{tab:moments}), indicating no detected difference at the widest validated window without establishing statistical equivalence or a physical crossover velocity.

To test which line-profile families are consistent with the $\PTJWSTPresetN$-source increase, the lower rows of Figure~\ref{fig:moments}(b) compare $\Delta s_{2500\to 4500}$ against synthetic profiles injected into the empirical noise of the exact same $\PTJWSTPresetN$ JWST spectra (Appendix~\ref{app:injections}). Single-Gaussian inputs anchored to each object's measured \FWHMsg\ have a noiseless finite-window median increment of $\PTMockPresetJWSTGaussDeltaTrue$~dex, and propagation through empirical noise and continuum fitting yields a 95\% conditional prediction interval of $[\PTMockPresetJWSTGaussDeltaLo, \PTMockPresetJWSTGaussDeltaHi]$~dex with median $\PTMockPresetJWSTGaussDeltaMed$~dex. Core-plus-wing inputs placing $5$--$35\%$ of the broad flux into a $3\times$ wider component have a noiseless median increment of $\PTMockPresetJWSTWingDeltaTrue$~dex and a 95\% conditional prediction interval of $[\PTMockPresetJWSTWingDeltaLo, \PTMockPresetJWSTWingDeltaHi]$~dex with median $\PTMockPresetJWSTWingDeltaMed$~dex; because pure Gaussians carry negligible flux beyond $2500~\kms$, outer-window noise perturbs a smaller baseline \sigV\ and inflates $\log_{10}\sigV$ more strongly than when a real wing is present, making the measurement noise itself profile-dependent. At the sample level, both conditional prediction intervals encompass the observed $\PTJWSTPresetDeltaSigned$~dex increase, as do the corresponding $\PTJWSTFullN$-source intervals in Table~\ref{tab:moments}, so current signal-to-noise ratios do not yet separate intrinsic wing diversity from finite-window truncation and wing noise. For SDSS, self-similar profiles anchored to \FWHMsg\ already predict larger noiseless median increments ($\PTMockPresetSDSSGaussDeltaTrue$ and $\PTMockPresetSDSSWingDeltaTrue$~dex, becoming $\PTMockPresetSDSSGaussDeltaLo$--$\PTMockPresetSDSSGaussDeltaHi$ and $\PTMockPresetSDSSWingDeltaLo$--$\PTMockPresetSDSSWingDeltaHi$~dex with noise) than the observed $\PTSDSSPresetDeltaSigned$~dex rise, reflecting the known anti-correlation between $\FWHM/\sigma$ and \FWHM\ in Type-1 AGNs \citep{Collin2006}; the SDSS spectra therefore serve as an empirical reference.

\subsection{Velocity-Dependent Balmer Weights}
\label{sec:res_balmer}

\begin{figure*}[t]
\centering
\includegraphics[width=\textwidth]{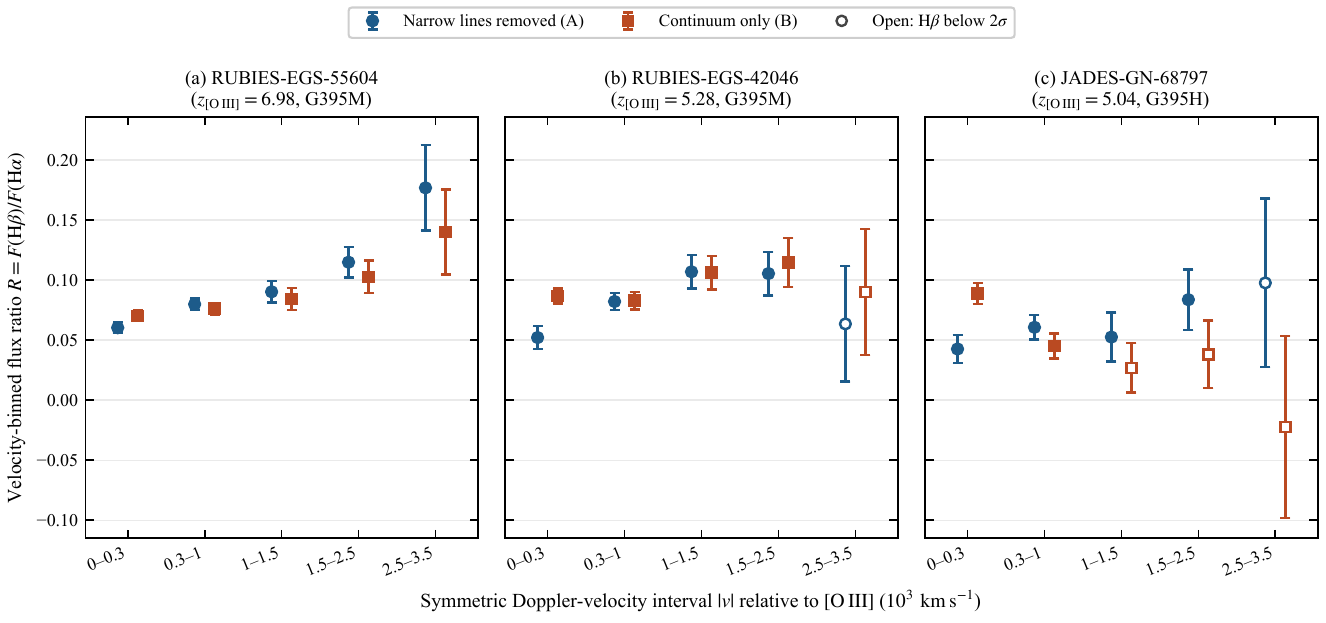}
\caption{\Hb\ becomes relatively stronger toward the wings in RUBIES-EGS-55604. Panels show velocity-binned flux ratios $R = F(\Hb)/F(\Ha)$ across five symmetric Doppler-velocity intervals relative to local systemic \OIII\ in the three highest-weight same-grating JWST LRDs: \textbf{(a)} RUBIES-EGS-55604, \textbf{(b)} RUBIES-EGS-42046, and \textbf{(c)} JADES-GN-68797. Blue circles show ratios after joint continuum and narrow-line subtraction (decomposition~A); terracotta squares show ratios after subtracting only a local continuum without narrow-line removal (decomposition~B). Both decompositions operate on the same combined grating spectrum, match \Ha\ to the \Hb\ velocity resolution, and retain Balmer absorption. Error bars indicate $1\sigma$ pixel uncertainties, and open symbols mark intervals with $\Hb < 2\sigma$, plotted as signed ratios without positivity clipping; full numerical values and decrements $D = 1/R$ appear in Appendix Table~\ref{tab:balmer_windows}.}
\label{fig:balmer}
\end{figure*}

Figure~\ref{fig:balmer} examines how \Hb\ and \Ha\ weight the same Doppler-velocity intervals in the three highest-weight dual-line LRDs, extending the center-to-wing Balmer decrement decline found by \citet[][Section~V.3.1 and Figure~9]{Chen2026} by comparing decompositions~A and~B on identical combined grating spectra (Appendix Table~\ref{tab:balmer_windows}). With \Ha\ matched to the \Hb\ velocity resolution in both cases, the two decompositions differ only in continuum placement and narrow-line removal, isolating the sensitivity of $R(v)$ to spectral decomposition.

In RUBIES-EGS-55604 (Figure~\ref{fig:balmer}a), which has the highest \Hb\ signal-to-noise ratio of the three targets, the continuum-only decrement $F(\Ha)/F(\Hb)$ declines from $\POneBSrcFiveFiveZeroThreeD$ at $0$--$300~\kms$ to $\POneBSrcFiveFiveOneFiveTwoFiveD$ at $1500$--$2500~\kms$ and $\POneBSrcFiveFiveTwoFiveThreeFiveD$ at $2500$--$3500~\kms$, corresponding to $R = \POneBSrcFiveFiveZeroThreeR \pm \POneBSrcFiveFiveZeroThreeRErr$ at line center and $\POneBSrcFiveFiveTwoFiveThreeFiveR \pm \POneBSrcFiveFiveTwoFiveThreeFiveRErr$ in the outer wing under decomposition~B. The same outward decline persists after narrow-line subtraction, with $D = \POneASrcFiveFiveZeroThreeD \to \POneASrcFiveFiveOneFiveTwoFiveD \to \POneASrcFiveFiveTwoFiveThreeFiveD$ under decomposition~A, showing that outer velocity intervals carry greater relative flux weight in \Hb\ than in \Ha\ under either spectral decomposition. By contrast, in RUBIES-EGS-42046 (Figure~\ref{fig:balmer}b) the central decrement is sensitive to narrow-line modeling, shifting from $D = \POneASrcFourTwoZeroThreeD$ in decomposition~A to $\POneBSrcFourTwoZeroThreeD$ in decomposition~B. Across $300$--$1000~\kms$, decomposition~A also shows a tentative blue-to-red asymmetry of $D = \POneASrcFourTwoThreeTenBlueD$ versus $\POneASrcFourTwoThreeTenRedD$ overlapping the blueshifted Balmer absorber reported in Table~2 of \citet{Chen2026}, while \Hb\ falls below $2\sigma$ at $2500$--$3500~\kms$. In JADES-GN-68797 (Figure~\ref{fig:balmer}c), continuum-only subtraction leaves \Hb\ below $2\sigma$ in every bin beyond $1000~\kms$.

Comparing these velocity-binned profiles with the $\POneTmplAllThirtyThreeN$-source inverse-variance-weighted template ratio of $\POneTmplAllThirtyThreeRatio$ (95\% conditional interval $[\POneTmplAllThirtyThreeLo, \POneTmplAllThirtyThreeHi]$; Appendix~\ref{app:template_weights}) clarifies why the two statistics differ. The template fit normalizes each object's broad-\Ha\ profile to \Hb\ across $300 \le |v| \le 1000~\kms$ relative to the fitted broad-\Ha\ centroid---which blueshifted absorption pulls redward of systemic \OIII, and within which narrow-line nuisance modes are projected out (Appendix~\ref{app:template_weights})---and measures an inverse-variance-weighted wing amplitude across $1500 \le |v| \le 4500~\kms$ dominated by the $1500$--$2500~\kms$ shoulder, whereas Figure~\ref{fig:balmer} integrates each velocity bin directly down to $0$--$300~\kms$ in the \OIII\ frame. A near-unity wing-to-shoulder template ratio across the $\POneTmplTopThreeN$ dominant objects, which supply $\POneTmplTopThreeWtPct\%$ of the total template weight, therefore coexists naturally with the center-to-wing gradient in RUBIES-EGS-55604, while the remaining $\POneTmplOtherThirtyN$ fainter targets carry $\POneTmplOtherThirtyWtPct\%$ of the weight and remain poorly constrained at $\POneTmplOtherThirtyRatio$ with a 95\% interval of $[\POneTmplOtherThirtyLo, \POneTmplOtherThirtyHi]$ spanning both zero and unity.

\subsection{Integrated-Width Population Context}
\label{sec:res_context}

\begin{figure*}[t]
\centering
\includegraphics[width=\textwidth]{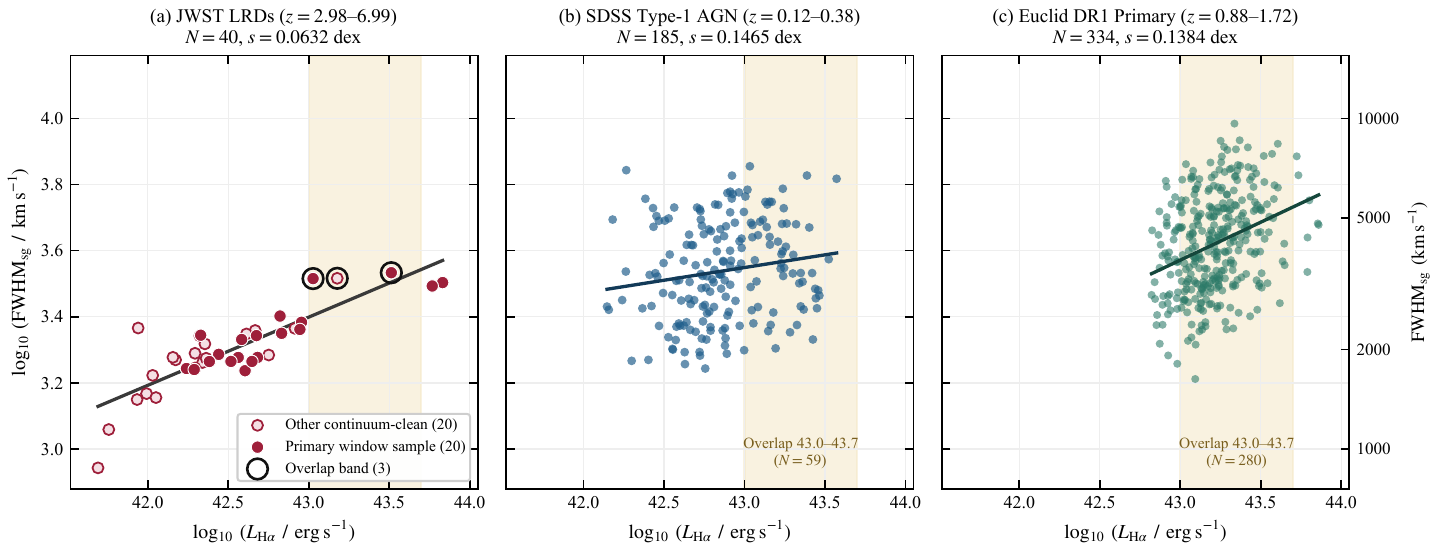}
\caption{Integrated broad-\Ha\ single-Gaussian widths (\FWHMsg) in three independently selected samples on common axes, with uniform OLS residual scatter $s$ at fixed observed \LHa\ listed above each panel. Gold shading marks the shared luminosity interval $\log_{10}(\LHa/\mathrm{erg\,s^{-1}}) = 43.0$--$43.7$; outer black rings in panel~\textbf{(a)} identify its $\JWSTBridgeN$ JWST sources. \textbf{(a)} JWST/NIRSpec LRDs ($N = \JWSTCleanN$ continuum-clean baseline, distinguishing the $\PTJWSTPresetN$ primary default-sideband window sources in filled crimson from the remaining $\JWSTOtherCleanN$ continuum-clean sources). \textbf{(b)} Quality-controlled SDSS DR7 Type-1 AGNs ($N = \SDSSCleanN$). \textbf{(c)} Euclid DR1 primary broad-\Ha\ sample ($N = \EuclidMainN$, $19.0 < H_{\mathrm{aper}} < \EuclidHLimit$).}
\label{fig:three_survey}
\end{figure*}

Figure~\ref{fig:three_survey} places integrated \FWHMsg\ at fixed observed \LHa\ on common axes across all three surveys using the identical OLS residual standard deviation. The $\JWSTCleanN$ continuum-clean JWST LRDs exhibit substantially smaller observed residual scatter about the integrated width--luminosity relation, with $s = \JWSTCleanSGScatterUniform$~dex across all $\JWSTCleanN$ objects and $\PTJWSTPresetSGScatter$~dex across the $\PTJWSTPresetN$ primary window targets, than either the $\SDSSCleanN$ SDSS Type-1 AGNs ($s = \SDSSCleanSGScatterUniform$~dex) or the $\EuclidMainN$ Euclid DR1 AGNs ($s = \EuclidMainSGScatterUniform$~dex). Independently, hierarchical modeling of the $\EuclidMainN$ Euclid DR1 AGNs conditioned on total rest-optical continuum luminosity and corrected for flux-limit selection yields an intrinsic width scatter of $\sigma_w = \EuclidMainSigmaW$~dex with a 95\% credible interval of $[\EuclidMainSigmaWLowerNinetyFive, \EuclidMainSigmaWUpperNinetyFive]$~dex. Because only $\JWSTBridgeN$ JWST sources fall inside the shared $\log_{10}(\LHa/\mathrm{erg\,s^{-1}}) = 43.0$--$43.7$ band alongside $\SDSSBridgeN$ SDSS and $\EuclidBridgeN$ Euclid AGNs, and a $\BridgeJWSTEuclidContinuumGap$~dex continuum luminosity gap separates the brightest JWST and faintest Euclid overlap objects, the three surveys provide a population reference for integrated \Ha\ widths rather than a test of redshift evolution at fixed luminosity.

\section{Discussion}
\label{sec:discussion}

Our two profile analyses probe complementary observables: population-level scatter in the \Ha\ finite-window line dispersion (Figure~\ref{fig:moments}) and single-object velocity-binned \Hb/\Ha\ ratios (Figure~\ref{fig:balmer}). Together, they clarify what an integrated LRD line width measures. Within the narrowest symmetric window ($V = 1500~\kms$), where flux density varies modestly across $[-V, +V]$, the finite-window line dispersion approaches $\sigV = V/\sqrt{3}$ (Appendix~\ref{app:exact_int}), compressing differences among profiles of distinct overall breadth. As $V$ widens from $2500$ to $4500~\kms$, outer pixels gain $(v - \bar{v}_V)^2$ weight and the observed scatter across the same $\PTJWSTPresetN$ LRDs more than doubles from $\PTJWSTPresetSTwoFive$ to $\PTJWSTPresetSFourFive$~dex. Because Gaussian and core-plus-wing injections into the empirical noise of those $\PTJWSTPresetN$ spectra both accommodate the observed $\PTJWSTPresetDeltaSigned$~dex rise, the present data establish how the measured scatter depends on velocity window without isolating a noise-free ``line-core'' invariant or separating intrinsic wing diversity from truncation and noise.

On the individual-source level, a uniform foreground dust screen attenuates $F(\Hb)/F(\Ha)$ by nearly the same factor across all Doppler velocities of a single line; while dust can shift the integrated Balmer decrement, producing a center-to-wing gradient requires velocity-dependent intrinsic emission or velocity-dependent radiative transfer. In RUBIES-EGS-55604, outer Doppler intervals carry a higher \Hb/\Ha\ flux ratio under either spectral decomposition, corroborating \citet{Chen2026}. Physically, higher-$|v|$ emission can arise in stratified broad-line gas where density and $n=2$ optical depth vary across emission zones \citep{OsterbrockFerland2006,InayoshiMaiolino2025,Chen2026,Madau2026b,Scholtz2026}, alongside velocity-selective Balmer absorption in dense gas \citep{deGraaffRuby2025,DEugenio2026,Matthee2026} or frequency redistribution and differential escape through optically thick envelopes \citep[][and the review by \citealt{InayoshiHo}]{Laor2006,Brazzini2025,Rusakov2026,Chang2025,NaiduBHStar2026,deGraaff2026}. Because scattering and absorption can decouple observed Doppler shift $v$ from bulk orbital speed, velocity-binned ratios alone neither map $v$ uniquely onto radius nor establish whether \Ha\ and \Hb\ yield discrepant black hole masses.

Viewed alongside the broader integrated-width distributions of SDSS and Euclid DR1 AGNs in Figure~\ref{fig:three_survey}, these findings set a direct forward-modeling requirement. Without independent black hole masses or reverberation lags, our spectra do not constrain the numerical value of a universal virial factor \fvir. Instead, any dynamical, optical-depth, or scattering model that uses a finite-window line dispersion or integrated width to infer LRD black hole masses or kinematics must evaluate its predictions over the \emph{same Balmer transition, width definition, and velocity interval} as the observations, reproducing both the window-dependent scatter $s(V)$ in Figure~\ref{fig:moments} after empirical-noise propagation and the velocity-binned Balmer ratios in bright dual-line targets such as RUBIES-EGS-55604. Deeper NIRSpec grating spectra will be needed to validate \Ha\ moments beyond $4500~\kms$, extend resolved \Hb/\Ha\ profiles to fainter LRDs, and test width-scaled windows once boundary uncertainties are propagated.

\section{Conclusions}
\label{sec:conclusions}

Holding the objects, width definition, and broad-\Ha\ luminosities fixed across velocity intervals yields two main empirical conclusions:
\begin{enumerate}
\item Across $\PTJWSTPresetN$ high-S/N JWST LRDs with default continuum sidebands, the observed scatter in broad-\Ha\ finite-window line dispersion \sigV\ at fixed \LHa\ rises from $\PTJWSTPresetSTwoFive$~dex at $V = 2500~\kms$ to $\PTJWSTPresetSFourFive$~dex at $V = 4500~\kms$, a paired increase of $\PTJWSTPresetDeltaSigned$~dex with 95\% bootstrap interval $[\PTJWSTPresetDeltaLo, \PTJWSTPresetDeltaHi]$~dex (Table~\ref{tab:moments}), versus $\PTSDSSPresetDeltaSigned$~dex ($[\PTSDSSPresetDeltaLo, \PTSDSSPresetDeltaHi]$~dex) in $\PTSDSSPresetNFourFive$ broader SDSS Type-1 AGNs and integrated-width scatters of $\JWSTCleanSGScatterUniform$, $\SDSSCleanSGScatterUniform$, and $\EuclidMainSGScatterUniform$~dex across JWST ($N = \JWSTCleanN$), SDSS ($N = \SDSSCleanN$), and Euclid DR1 ($N = \EuclidMainN$). Synthetic profile injections into the empirical noise of the same $\PTJWSTPresetN$ JWST spectra show that single-Gaussian and core-plus-wing models both accommodate the observed $\Delta s_{2500\to 4500}$ at the sample level, leaving intrinsic wing diversity, truncation, and wing noise unseparated.
\item In RUBIES-EGS-55604, outer velocity intervals carry greater relative \Hb\ flux than \Ha\ under both narrow-line-subtracted and continuum-only decompositions of the same combined grating spectrum, with $F(\Ha)/F(\Hb)$ declining from $\POneBSrcFiveFiveZeroThreeD$ to $\POneBSrcFiveFiveOneFiveTwoFiveD$ and $\POneBSrcFiveFiveTwoFiveThreeFiveD$ from center to wing after continuum subtraction alone. Comparisons between LRD broad-line models and observations must therefore match both the Balmer transition and the velocity interval included in the finite-window line dispersion.
\end{enumerate}

\section*{Acknowledgments}
The author is grateful to his parents for their constant support. This research received no external grant funding.

Some of the data products presented herein were retrieved from the Dawn JWST Archive (DJA). DJA is an initiative of the Cosmic Dawn Center (DAWN), which is funded by the Danish National Research Foundation under grant DNRF140. This work is based on observations made with the NASA/ESA/CSA James Webb Space Telescope. The reduced spectra used here were retrieved from the DAWN JWST Archive; the underlying JWST observations are archived in the Mikulski Archive for Space Telescopes at the Space Telescope Science Institute, which is operated by the Association of Universities for Research in Astronomy, Inc., under NASA contract NAS5-03127 for JWST. These observations are associated with programs \#1181 (PI D.~J. Eisenstein), \#1210 (PI N. L\"utzgendorf), \#1286 (PI N. L\"utzgendorf), \#1345 (PI S.~L. Finkelstein), \#4106 (PI E. Nelson), \#4233 (PI A. de Graaff), \#5718 (PI D.~D. Kocevski), and \#8047 (PI B. Wang). The author acknowledges the teams of programs \#1345 and \#4233 for developing their observing programs with zero-exclusive-access periods.

Funding for the SDSS and SDSS-II has been provided by the Alfred P. Sloan Foundation, the Participating Institutions, the National Science Foundation, the U.S. Department of Energy, the National Aeronautics and Space Administration, the Japanese Monbukagakusho, the Max Planck Society, and the Higher Education Funding Council for England. The SDSS Web Site is \url{http://www.sdss.org/}. The SDSS is managed by the Astrophysical Research Consortium for the Participating Institutions.

The Participating Institutions are the American Museum of Natural History, Astrophysical Institute Potsdam, University of Basel, University of Cambridge, Case Western Reserve University, University of Chicago, Drexel University, Fermilab, the Institute for Advanced Study, the Japan Participation Group, Johns Hopkins University, the Joint Institute for Nuclear Astrophysics, the Kavli Institute for Particle Astrophysics and Cosmology, the Korean Scientist Group, the Chinese Academy of Sciences (LAMOST), Los Alamos National Laboratory, the Max-Planck-Institute for Astronomy (MPIA), the Max-Planck-Institute for Astrophysics (MPA), New Mexico State University, Ohio State University, University of Pittsburgh, University of Portsmouth, Princeton University, the United States Naval Observatory, and the University of Washington.

This work has made use of the Euclid DR1 data from the \emph{Euclid} mission of the European Space Agency (ESA), 2026, \url{https://doi.org/10.57780/esa-d079c51}, processed by the \emph{Euclid} Consortium (\url{https://www.euclid-ec.org/}).

Reproduction scripts, derived catalogs, and figure-generation code for this study are archived at \url{https://github.com/NianSi-AI/lrd-balmer-velocity-windows} and on Zenodo at \dataset[doi:10.5281/zenodo.24426322]{https://doi.org/10.5281/zenodo.24426322}.

AI disclosure: Gemini (Google), Claude (Anthropic), Codex (OpenAI), and DeepSeek were used under the author's direction to audit physical and statistical arguments, write and execute analysis and Matplotlib figure-generation code, cross-check literature citations, and refine prose from the author's drafts. All text, numbers, figures, and references were verified by the author, who bears full responsibility for the content.

\section*{Author Contributions}
H.W. conceived and designed the study, performed the data analysis and modeling, and wrote the manuscript.

\facilities{JWST (NIRSpec), Sloan, Euclid (NISP, VIS)}

\software{\texttt{msaexp} \citep{msaexp}, NumPy \citep{harris2020array}, SciPy \citep{2020SciPy-NMeth}, Matplotlib \citep{Hunter:2007}, Astropy \citep{astropy:2013,astropy:2018,astropy:2022}}

\clearpage
\appendix
\restartappendixnumbering
\makeatletter
\@addtoreset{figure}{section}
\@addtoreset{table}{section}
\makeatother
\renewcommand{\theHfigure}{\Alph{section}\arabic{figure}}
\renewcommand{\theHtable}{\Alph{section}\arabic{table}}
\twocolumngrid

\section{Finite-Window Integration, Recovery Calibration, and Near-Line Sidebands}
\label{app:calibration}

\begin{figure*}[t]
\centering
\includegraphics[width=0.92\textwidth]{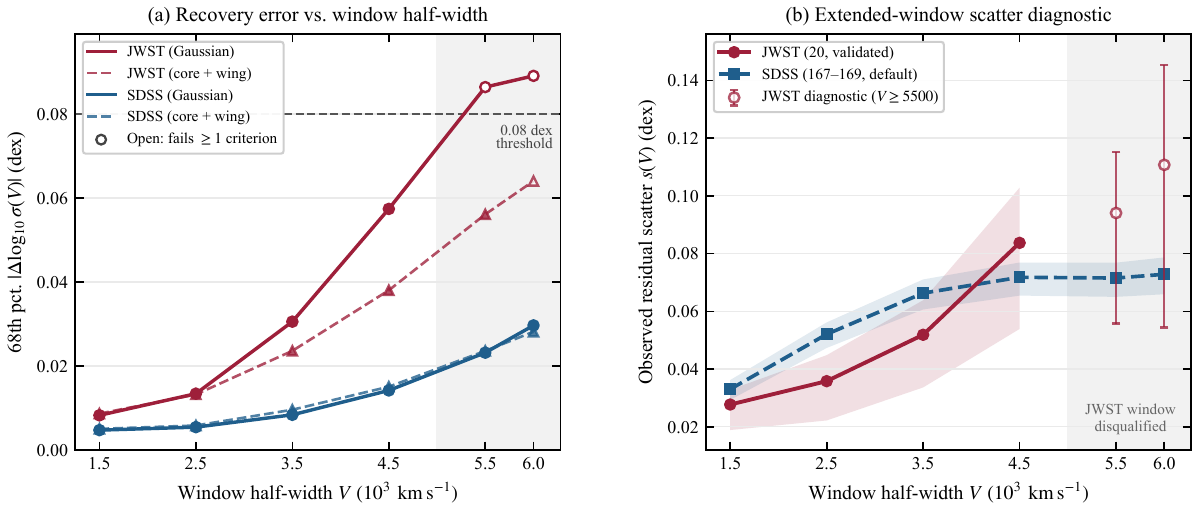}
\caption{Recovery calibration and extended-window diagnostics for the \Ha\ finite-window line dispersion. \textbf{(a)} 68th-percentile $|\Delta\log_{10}\sigV|$ across $V = 1500$--$6000~\kms$. Open symbols fail at least one per-profile qualification threshold ($\ge 85\%$ valid recoveries, $|\text{median log bias}| \le 0.03$~dex, or 68th-percentile $|\Delta\log_{10}\sigV| \le 0.08$~dex; the dashed horizontal line marks only the $0.08$~dex error limit, so JWST core-plus-wing at $V = 6000~\kms$ is open owing to its $\PTCalJWSTWingSixZeroUsablePct\%$ recovery rate). \textbf{(b)} Extended-window observed scatter diagnostic showing validated $V \le 4500~\kms$ curves alongside disqualified JWST windows at $V = 5500$ and $6000~\kms$ (open circles).}
\label{fig:app_cal}
\end{figure*}

\subsection{Positive-Overlap Pixel Quadrature and Flat-Window Limit}
\label{app:exact_int}

On each native wavelength grid $\{\lambda_i\}$, midpoint edges $\lambda_{i,\pm} = \tfrac{1}{2}(\lambda_i + \lambda_{i\pm 1})$, extrapolated symmetrically at spectrum endpoints, convert to rest-frame velocity edges $[v_{i,-}, v_{i,+}]$. In the \Ha\ moment scan, each pixel edge is further restricted to lie within $0.55$ times the median pixel spacing of $v_i$ so that masked or missing wavelength intervals are never bridged. Intersecting each valid pixel with $[v_{\mathrm{lo}}, v_{\mathrm{hi}}]$ and retaining only positive overlaps via $a_i = \max(v_{i,-}, v_{\mathrm{lo}})$ and $b_i = \max(a_i, \min(v_{i,+}, v_{\mathrm{hi}}))$ gives exact piecewise-constant integrals for window flux $F(V) = \sum_i f_i (b_i - a_i)$, centroid $\bar{v}_V = [2 F(V)]^{-1}\sum_i f_i (b_i^2 - a_i^2)$, and finite-window line dispersion
\begin{equation}
\sigV = \left[ \frac{1}{3 F(V)} \sum_i f_i \left( (b_i - \bar{v}_V)^3 - (a_i - \bar{v}_V)^3 \right) \right]^{1/2}.
\label{eq:app_sigma_v}
\end{equation}
For constant flux $f_i = f_0$ across $[-V, +V]$, Equation~(\ref{eq:app_sigma_v}) yields $F(V) = 2 f_0 V$, $\bar{v}_V = 0$, and $\sigV = V/\sqrt{3}$. Because finite $V$ truncates broader lines more strongly than narrower lines, the noiseless $\PTJWSTPresetN$-source Gaussian mock scatter at $V = 2500~\kms$ ($\PTMockPresetJWSTGaussTrueTwoFive$~dex for JWST and $\PTMockPresetSDSSGaussTrueTwoFive$~dex for SDSS) lies below its untruncated full-profile counterpart of $\PTMockPresetJWSTGaussTrueFull$ and $\PTMockPresetSDSSGaussTrueFull$~dex, respectively.

\subsection{Window Recovery Calibration and Empirical-Noise Mock Ensembles}
\label{app:injections}

\begin{deluxetable*}{lcccccc}
\tabletypesize{\scriptsize}
\tabcolsep=4.5pt
\tablewidth{0pt}
\tablecaption{Injection-Recovery Window Calibration Across Tested Velocity Windows\label{tab:calibration}}
\tablehead{
\colhead{Window $V$} &
\multicolumn{3}{c}{Gaussian Input} &
\multicolumn{3}{c}{Core + Wing Input ($5$--$35\%$ Wing Flux)} \\
\cmidrule(lr){2-4} \cmidrule(lr){5-7}
\colhead{(\kms)} &
\colhead{Recovered (\%)} &
\colhead{Med.\ Bias (dex)} &
\colhead{68th Err.\ (dex)} &
\colhead{Recovered (\%)} &
\colhead{Med.\ Bias (dex)} &
\colhead{68th Err.\ (dex)}
}
\startdata
\cutinhead{JWST, $\PTCalJWSTAttemptedTargetsN$ targets attempted (default sidebands)}
1500 & $\PTCalJWSTGaussOneFiveUsablePct$ & $\PTCalJWSTGaussOneFiveBias$ & $\PTCalJWSTGaussOneFiveErrSixEight$ & $\PTCalJWSTWingOneFiveUsablePct$ & $\PTCalJWSTWingOneFiveBias$ & $\PTCalJWSTWingOneFiveErrSixEight$ \\
2500 & $\PTCalJWSTGaussTwoFiveUsablePct$ & $\PTCalJWSTGaussTwoFiveBias$ & $\PTCalJWSTGaussTwoFiveErrSixEight$ & $\PTCalJWSTWingTwoFiveUsablePct$ & $\PTCalJWSTWingTwoFiveBias$ & $\PTCalJWSTWingTwoFiveErrSixEight$ \\
3500 & $\PTCalJWSTGaussThreeFiveUsablePct$ & $\PTCalJWSTGaussThreeFiveBias$ & $\PTCalJWSTGaussThreeFiveErrSixEight$ & $\PTCalJWSTWingThreeFiveUsablePct$ & $\PTCalJWSTWingThreeFiveBias$ & $\PTCalJWSTWingThreeFiveErrSixEight$ \\
4500 & $\PTCalJWSTGaussFourFiveUsablePct$ & $\PTCalJWSTGaussFourFiveBias$ & $\PTCalJWSTGaussFourFiveErrSixEight$ & $\PTCalJWSTWingFourFiveUsablePct$ & $\PTCalJWSTWingFourFiveBias$ & $\PTCalJWSTWingFourFiveErrSixEight$ \\
5500\tablenotemark{a} & $\PTCalJWSTGaussFiveFiveUsablePct^{\ast}$ & $\PTCalJWSTGaussFiveFiveBias$ & $\PTCalJWSTGaussFiveFiveErrSixEight^{\ast}$ & $\PTCalJWSTWingFiveFiveUsablePct$ & $\PTCalJWSTWingFiveFiveBias$ & $\PTCalJWSTWingFiveFiveErrSixEight$ \\
6000\tablenotemark{a} & $\PTCalJWSTGaussSixZeroUsablePct^{\ast}$ & $\PTCalJWSTGaussSixZeroBias$ & $\PTCalJWSTGaussSixZeroErrSixEight^{\ast}$ & $\PTCalJWSTWingSixZeroUsablePct^{\ast}$ & $\PTCalJWSTWingSixZeroBias$ & $\PTCalJWSTWingSixZeroErrSixEight$ \\
\cutinhead{SDSS, $\PTCalSDSSAttemptedTargetsN$ targets attempted (default sidebands)}
1500 & $\PTCalSDSSGaussOneFiveUsablePct$ & $\PTCalSDSSGaussOneFiveBias$ & $\PTCalSDSSGaussOneFiveErrSixEight$ & $\PTCalSDSSWingOneFiveUsablePct$ & $\PTCalSDSSWingOneFiveBias$ & $\PTCalSDSSWingOneFiveErrSixEight$ \\
2500 & $\PTCalSDSSGaussTwoFiveUsablePct$ & $\PTCalSDSSGaussTwoFiveBias$ & $\PTCalSDSSGaussTwoFiveErrSixEight$ & $\PTCalSDSSWingTwoFiveUsablePct$ & $\PTCalSDSSWingTwoFiveBias$ & $\PTCalSDSSWingTwoFiveErrSixEight$ \\
3500 & $\PTCalSDSSGaussThreeFiveUsablePct$ & $\PTCalSDSSGaussThreeFiveBias$ & $\PTCalSDSSGaussThreeFiveErrSixEight$ & $\PTCalSDSSWingThreeFiveUsablePct$ & $\PTCalSDSSWingThreeFiveBias$ & $\PTCalSDSSWingThreeFiveErrSixEight$ \\
4500 & $\PTCalSDSSGaussFourFiveUsablePct$ & $\PTCalSDSSGaussFourFiveBias$ & $\PTCalSDSSGaussFourFiveErrSixEight$ & $\PTCalSDSSWingFourFiveUsablePct$ & $\PTCalSDSSWingFourFiveBias$ & $\PTCalSDSSWingFourFiveErrSixEight$ \\
5500 & $\PTCalSDSSGaussFiveFiveUsablePct$ & $\PTCalSDSSGaussFiveFiveBias$ & $\PTCalSDSSGaussFiveFiveErrSixEight$ & $\PTCalSDSSWingFiveFiveUsablePct$ & $\PTCalSDSSWingFiveFiveBias$ & $\PTCalSDSSWingFiveFiveErrSixEight$ \\
6000 & $\PTCalSDSSGaussSixZeroUsablePct$ & $\PTCalSDSSGaussSixZeroBias$ & $\PTCalSDSSGaussSixZeroErrSixEight$ & $\PTCalSDSSWingSixZeroUsablePct$ & $\PTCalSDSSWingSixZeroBias$ & $\PTCalSDSSWingSixZeroErrSixEight$ \\
\enddata
\tablecomments{Calibration tests injected $10$ empirical four-pixel block-noise realizations per attempted target and profile class under default sidebands, so recovered percentages divide valid recoveries by all attempted realizations in each cell (for example, $\PTCalJWSTGaussOneFiveOkTrials/\PTCalJWSTGaussOneFiveTotTrials = \PTCalJWSTGaussOneFiveUsablePct\%$ for JWST and $\PTCalSDSSGaussOneFiveOkTrials/\PTCalSDSSGaussOneFiveTotTrials = \PTCalSDSSGaussOneFiveUsablePct\%$ for SDSS at $V = 1500~\kms$ Gaussian), including spectra that lack default sidebands. Joint pooled window qualification requires both profile classes to achieve $\ge 85\%$ valid recoveries, $|\text{median log bias}| \le 0.03$~dex, and 68th-percentile $|\Delta\log_{10}\sigV| \le 0.08$~dex across the attempted ensemble; asterisks ($^{\ast}$) mark individual threshold violations.}
\tablenotetext{a}{Loses joint JWST window qualification: $V = 5500~\kms$ fails the Gaussian recovery rate and error thresholds, and $V = 6000~\kms$ fails the Gaussian recovery rate and error thresholds as well as the core-plus-wing recovery rate. Across all $\JWSTCleanN$ continuum-clean JWST targets ($\PTJWSTFortyNFourFive$ valid at $4500~\kms$, with observed $s = \PTJWSTFortySTwoFive \to \PTJWSTFortySFourFive$~dex), Gaussian recovery already falls short at $V = 4500~\kms$ ($\PTCalJWSTFortyGaussFourFiveUsablePct\%$ recovered, $\PTCalJWSTFortyGaussFourFiveErrSixEight$~dex 68th-percentile error).}
\end{deluxetable*}

We established the pooled ensemble window qualification thresholds prior to measuring the observed finite-window moments (Figure~\ref{fig:app_cal}; Table~\ref{tab:calibration}) by injecting $10$ four-pixel block-noise realizations per attempted target across $\PTCalJWSTAttemptedTargetsN$ high-S/N JWST and $\PTCalSDSSAttemptedTargetsN$ SDSS spectra, including synthetic broad \Ha, narrow \Ha+\NII, linear continua, and blueshifted absorption in a subset of realizations. Separately, to construct the conditional prediction intervals for $\Delta s_{2500\to 4500}$, we generated $60$ empirical-noise realizations per JWST source and $12$ per SDSS source for each profile class, fixing each object's intrinsic core FWHM to its measured \FWHMsg\ and including blueshifted absorption in a subset of realizations prior to LSF convolution. Drawing one realization recovered in both windows per source preserves paired pixel noise across $4000$ synthetic sample draws, whose equal-tailed 2.5th--97.5th percentiles give the $\PTJWSTPresetN$-source default-sideband prediction intervals in Figure~\ref{fig:moments}(b) and the $\PTJWSTFullN$-source intervals in Table~\ref{tab:moments}.

\subsection{Near-Line Continuum Sidebands for Detector-Edge Spectra}
\label{app:edge}

Following the primary default-sideband analysis, we tested closer continuum windows at $[+5600, +6300]~\kms$ on the red side and $[-6300, -5400]~\kms$ on the blue side for spectra truncated near detector edges: RUBIES-EGS-50052 and JADES-GS-38562 in JWST, together with the additional SDSS spectra entering the $\PTSDSSFullNFourFive$--$\PTSDSSFullNOneFive$ set in Table~\ref{tab:moments}. At $V = 4500~\kms$, RUBIES-EGS-50052 achieves $\PTEdgeFiveZeroZeroFiveTwoGaussUsablePct\%$ valid recoveries with a $\PTEdgeFiveZeroZeroFiveTwoGaussErrSixEight$~dex 68th-percentile error for Gaussian inputs and $\PTEdgeFiveZeroZeroFiveTwoWingUsablePct\%$ with a $\PTEdgeFiveZeroZeroFiveTwoWingErrSixEight$~dex error for core-plus-wing inputs. By contrast, JADES-GS-38562 is noisier near the detector cutoff, reaching $\PTEdgeThreeEightFiveSixTwoGaussUsablePct\%$ valid recoveries and $\PTEdgeThreeEightFiveSixTwoGaussErrSixEight$~dex error for Gaussian inputs and $\PTEdgeThreeEightFiveSixTwoWingUsablePct\%$ and $\PTEdgeThreeEightFiveSixTwoWingErrSixEight$~dex for core-plus-wing inputs, motivating the $\PTJWSTPresetN$ default-sideband objects as our primary sample.

\section{Velocity-Binned Balmer Ratios and Template Weight Breakdown}
\label{app:template_weights}

\begin{figure*}[t]
\centering
\includegraphics[width=0.52\textwidth]{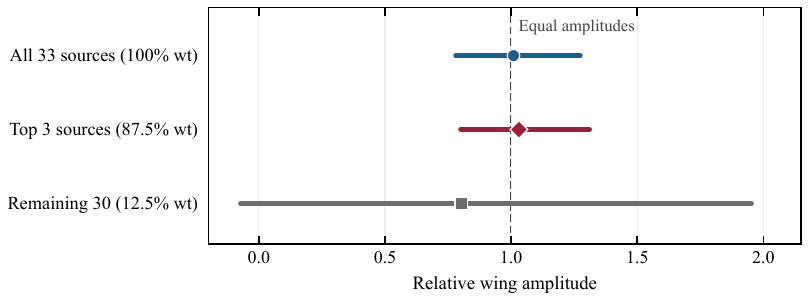}
\caption{\Hb\ template amplitude in the $1500$--$4500~\kms$ wing band relative to the $300$--$1000~\kms$ normalization band, with 95\% pixel-noise intervals conditional on fixed \Ha\ profiles, across all $\POneTmplAllThirtyThreeN$ same-grating sources ($\POneTmplAllThirtyThreeRatio$, $[\POneTmplAllThirtyThreeLo, \POneTmplAllThirtyThreeHi]$), the top $\POneTmplTopThreeN$ sources shown in Figure~\ref{fig:balmer} ($\POneTmplTopThreeWtPct\%$ weight; $\POneTmplTopThreeRatio$, $[\POneTmplTopThreeLo, \POneTmplTopThreeHi]$), and the remaining $\POneTmplOtherThirtyN$ sources ($\POneTmplOtherThirtyWtPct\%$ weight; $\POneTmplOtherThirtyRatio$, $[\POneTmplOtherThirtyLo, \POneTmplOtherThirtyHi]$).}
\label{fig:app_template}
\end{figure*}

Table~\ref{tab:balmer_windows} lists the five-interval $R = F(\Hb)/F(\Ha)$ and $D = 1/R$ measurements shown in Figure~\ref{fig:balmer}. Figure~\ref{fig:app_template} shows how individual targets weight the $\POneTmplAllThirtyThreeN$-source same-grating template ratio discussed in Section~\ref{sec:res_balmer}, projecting out $150$ and $500~\kms$ Gaussian nuisance modes across $300 \le |v| \le 1000~\kms$ and using inverse-variance weights calibrated on $4096$ empirical block-noise realizations per source and evaluated on an independent set of $4096$ realizations. Because inverse-variance weights scale quadratically with line S/N, RUBIES-EGS-55604, RUBIES-EGS-42046, and JADES-GN-68797 supply $\POneTmplTopThreeWtPct\%$ of the total weight, yielding a wing-to-shoulder ratio of $\POneTmplTopThreeRatio$ with a 95\% interval of $[\POneTmplTopThreeLo, \POneTmplTopThreeHi]$. The remaining $\POneTmplOtherThirtyN$ fainter objects carry $\POneTmplOtherThirtyWtPct\%$ of the weight and give $\POneTmplOtherThirtyRatio$ with a 95\% interval of $[\POneTmplOtherThirtyLo, \POneTmplOtherThirtyHi]$, spanning both zero and unity because noisy \Hb\ amplitudes are retained without positivity clipping.

\begin{deluxetable*}{lcccccc}
\tabletypesize{\scriptsize}
\tablewidth{0pt}
\tablecaption{Velocity-Binned \Hb/\Ha\ Flux Ratios $R$ and Equivalent Decrements $D$ in Three High-Weight Same-Grating LRDs\label{tab:balmer_windows}}
\tablehead{
\colhead{$|v|$ Interval} &
\multicolumn{2}{c}{RUBIES-EGS-55604} &
\multicolumn{2}{c}{RUBIES-EGS-42046} &
\multicolumn{2}{c}{JADES-GN-68797} \\
\cmidrule(lr){2-3} \cmidrule(lr){4-5} \cmidrule(lr){6-7}
\colhead{(\kms)} &
\colhead{Narrow sub.\ (A)} &
\colhead{Cont.\ only (B)} &
\colhead{Narrow sub.\ (A)} &
\colhead{Cont.\ only (B)} &
\colhead{Narrow sub.\ (A)} &
\colhead{Cont.\ only (B)}
}
\startdata
$0$--$300$ &
  $\POneASrcFiveFiveZeroThreeR \pm \POneASrcFiveFiveZeroThreeRErr$ ($\POneASrcFiveFiveZeroThreeD$) &
  $\POneBSrcFiveFiveZeroThreeR \pm \POneBSrcFiveFiveZeroThreeRErr$ ($\POneBSrcFiveFiveZeroThreeD$) &
  $\POneASrcFourTwoZeroThreeR \pm \POneASrcFourTwoZeroThreeRErr$ ($\POneASrcFourTwoZeroThreeD$) &
  $\POneBSrcFourTwoZeroThreeR \pm \POneBSrcFourTwoZeroThreeRErr$ ($\POneBSrcFourTwoZeroThreeD$) &
  $\POneASrcSixEightZeroThreeR \pm \POneASrcSixEightZeroThreeRErr$ ($\POneASrcSixEightZeroThreeD$) &
  $\POneBSrcSixEightZeroThreeR \pm \POneBSrcSixEightZeroThreeRErr$ ($\POneBSrcSixEightZeroThreeD$) \\
$300$--$1000$ &
  $\POneASrcFiveFiveThreeTenR \pm \POneASrcFiveFiveThreeTenRErr$ ($\POneASrcFiveFiveThreeTenD$) &
  $\POneBSrcFiveFiveThreeTenR \pm \POneBSrcFiveFiveThreeTenRErr$ ($\POneBSrcFiveFiveThreeTenD$) &
  $\POneASrcFourTwoThreeTenR \pm \POneASrcFourTwoThreeTenRErr$ ($\POneASrcFourTwoThreeTenD$) &
  $\POneBSrcFourTwoThreeTenR \pm \POneBSrcFourTwoThreeTenRErr$ ($\POneBSrcFourTwoThreeTenD$) &
  $\POneASrcSixEightThreeTenR \pm \POneASrcSixEightThreeTenRErr$ ($\POneASrcSixEightThreeTenD$) &
  $\POneBSrcSixEightThreeTenR \pm \POneBSrcSixEightThreeTenRErr$ ($\POneBSrcSixEightThreeTenD$) \\
$1000$--$1500$ &
  $\POneASrcFiveFiveTenOneFiveR \pm \POneASrcFiveFiveTenOneFiveRErr$ ($\POneASrcFiveFiveTenOneFiveD$) &
  $\POneBSrcFiveFiveTenOneFiveR \pm \POneBSrcFiveFiveTenOneFiveRErr$ ($\POneBSrcFiveFiveTenOneFiveD$) &
  $\POneASrcFourTwoTenOneFiveR \pm \POneASrcFourTwoTenOneFiveRErr$ ($\POneASrcFourTwoTenOneFiveD$) &
  $\POneBSrcFourTwoTenOneFiveR \pm \POneBSrcFourTwoTenOneFiveRErr$ ($\POneBSrcFourTwoTenOneFiveD$) &
  $\POneASrcSixEightTenOneFiveR \pm \POneASrcSixEightTenOneFiveRErr$ ($\POneASrcSixEightTenOneFiveD$) &
  $\POneBSrcSixEightTenOneFiveR \pm \POneBSrcSixEightTenOneFiveRErr$ ($<2\sigma$) \\
$1500$--$2500$ &
  $\POneASrcFiveFiveOneFiveTwoFiveR \pm \POneASrcFiveFiveOneFiveTwoFiveRErr$ ($\POneASrcFiveFiveOneFiveTwoFiveD$) &
  $\POneBSrcFiveFiveOneFiveTwoFiveR \pm \POneBSrcFiveFiveOneFiveTwoFiveRErr$ ($\POneBSrcFiveFiveOneFiveTwoFiveD$) &
  $\POneASrcFourTwoOneFiveTwoFiveR \pm \POneASrcFourTwoOneFiveTwoFiveRErr$ ($\POneASrcFourTwoOneFiveTwoFiveD$) &
  $\POneBSrcFourTwoOneFiveTwoFiveR \pm \POneBSrcFourTwoOneFiveTwoFiveRErr$ ($\POneBSrcFourTwoOneFiveTwoFiveD$) &
  $\POneASrcSixEightOneFiveTwoFiveR \pm \POneASrcSixEightOneFiveTwoFiveRErr$ ($\POneASrcSixEightOneFiveTwoFiveD$) &
  $\POneBSrcSixEightOneFiveTwoFiveR \pm \POneBSrcSixEightOneFiveTwoFiveRErr$ ($<2\sigma$) \\
$2500$--$3500$ &
  $\POneASrcFiveFiveTwoFiveThreeFiveR \pm \POneASrcFiveFiveTwoFiveThreeFiveRErr$ ($\POneASrcFiveFiveTwoFiveThreeFiveD$) &
  $\POneBSrcFiveFiveTwoFiveThreeFiveR \pm \POneBSrcFiveFiveTwoFiveThreeFiveRErr$ ($\POneBSrcFiveFiveTwoFiveThreeFiveD$) &
  $\POneASrcFourTwoTwoFiveThreeFiveR \pm \POneASrcFourTwoTwoFiveThreeFiveRErr$ ($<2\sigma$) &
  $\POneBSrcFourTwoTwoFiveThreeFiveR \pm \POneBSrcFourTwoTwoFiveThreeFiveRErr$ ($<2\sigma$) &
  $\POneASrcSixEightTwoFiveThreeFiveR \pm \POneASrcSixEightTwoFiveThreeFiveRErr$ ($<2\sigma$) &
  $\POneBSrcSixEightTwoFiveThreeFiveR \pm \POneBSrcSixEightTwoFiveThreeFiveRErr$ ($<2\sigma$) \\
\enddata
\tablecomments{Entries list $R \equiv F(\Hb)/F(\Ha) \pm 1\sigma$ integrated over symmetric native-pixel overlaps relative to local systemic \OIII\ with \Ha\ matched to the \Hb\ velocity resolution, propagating matched-residual and pixel uncertainties in decomposition~A and pixel plus local continuum-fit covariance in decomposition~B. Equivalent decrements $D = 1/R$ appear in parentheses only where integrated $\Hb \ge 2\sigma$.}
\end{deluxetable*}

\onecolumngrid
\clearpage
\raggedbottom
\twocolumngrid
\bibliography{references}

@article{Wang2026,
  author        = {{Wang}, Haoyu},
  title         = {A Tight Broad-line Width--Luminosity Relation in Little Red Dots and Its Consequence for Virial Black Hole Masses},
  journal       = {arXiv e-prints},
  year          = 2026,
  eid           = {arXiv:2611.24217},
  pages         = {arXiv:2611.24217},
  doi           = {10.48550/arXiv.2611.24217},
  eprint        = {2611.24217},
  archivePrefix = {arXiv},
  primaryClass  = {astro-ph.GA}
}

@article{Chen2026,
  author        = {{Chen}, Chang-Hao and {Shangguan}, Jinyi and {Ho}, Luis C. and {Zhang}, Zijian and {Inayoshi}, Kohei and {Li}, Ruancun},
  title         = {{ABCD}: The Nuclear Structure of the Little Red Dots Revealed through Absorption, Break, Continuum, and Decrement},
  journal       = {arXiv e-prints},
  year          = 2026,
  eid           = {arXiv:2606.04711},
  pages         = {arXiv:2606.04711},
  doi           = {10.48550/arXiv.2606.04711},
  eprint        = {2606.04711},
  archivePrefix = {arXiv},
  primaryClass  = {astro-ph.GA}
}

@article{Yanagisawa2026,
  author        = {{Yanagisawa}, Hiroto and {Ouchi}, Masami and {Kiyota}, Tomokazu and {Kageura}, Yuta and {Ando}, Makoto and {Harikane}, Yuichi and {Nakane}, Minami and {Ono}, Yoshiaki and {Takeda}, Yui},
  title         = {{ATLAS. I.} A Scaling Relation of {LRDs} between Broad {H$\alpha$} and Bolometric Luminosities: Enhanced Broad {H$\alpha$} Emission Relative to Low-$z$ Type 1 {AGN}},
  journal       = {arXiv e-prints},
  year          = 2026,
  eid           = {arXiv:2606.09726},
  pages         = {arXiv:2606.09726},
  doi           = {10.48550/arXiv.2606.09726},
  eprint        = {2606.09726},
  archivePrefix = {arXiv},
  primaryClass  = {astro-ph.GA}
}

@article{Denney2009,
  author        = {{Denney}, Kelly D. and {Peterson}, Bradley M. and {Dietrich}, Matthias and {Vestergaard}, Marianne and {Bentz}, Misty C.},
  title         = {Systematic Uncertainties in Black Hole Masses Determined from Single-Epoch Spectra},
  journal       = {\apj},
  year          = 2009,
  volume        = {692},
  number        = {1},
  pages         = {246--264},
  doi           = {10.1088/0004-637X/692/1/246},
  eprint        = {0810.3234},
  archivePrefix = {arXiv},
  primaryClass  = {astro-ph}
}

@article{Matthee2024,
  author        = {{Matthee}, Jorryt and {Naidu}, Rohan P. and {Brammer}, Gabriel and {Chisholm}, John and {Eilers}, Anna-Christina and {Goulding}, Andy and {Greene}, Jenny and {Kashino}, Daichi and {Labb{\'e}}, Ivo and {Lilly}, Simon J. and {Mackenzie}, Ruari and {Oesch}, Pascal A. and {Weibel}, Andrea and {Wuyts}, Stijn and {Xiao}, Mengyuan and {Bordoloi}, Rongmon and {Bouwens}, Rychard and {van Dokkum}, Pieter and {Illingworth}, Garth and {Kramarenko}, Ivan and {Maseda}, Michael V. and {Mason}, Charlotte and {Meyer}, Romain A. and {Nelson}, Erica J. and {Reddy}, Naveen A. and {Shivaei}, Irene and {Simcoe}, Robert A. and {Yue}, Minghao},
  title         = {Little Red Dots: An Abundant Population of Faint Active Galactic Nuclei at $z \sim 5$ Revealed by the {EIGER} and {FRESCO} {JWST} Surveys},
  journal       = {\apj},
  year          = 2024,
  volume        = {963},
  number        = {2},
  pages         = {129},
  doi           = {10.3847/1538-4357/ad2345},
  eprint        = {2306.05448},
  archivePrefix = {arXiv},
  primaryClass  = {astro-ph.GA}
}

@article{Greene2024,
  author        = {{Greene}, Jenny E. and {Labb{\'e}}, Ivo and {Goulding}, Andy D. and {Furtak}, Lukas J. and {Chemerynska}, Iryna and {Kokorev}, Vasily and {Dayal}, Pratika and {Volonteri}, Marta and {Williams}, Christina C. and {Wang}, Bingjie and {Setton}, David J. and {Burgasser}, Adam J. and {Bezanson}, Rachel and {Atek}, Hakim and {Brammer}, Gabriel and {Cutler}, Sam E. and {Feldmann}, Robert and {Fujimoto}, Seiji and {Glazebrook}, Karl and {de Graaff}, Anna and {Khullar}, Gourav and {Leja}, Joel and {Marchesini}, Danilo and {Maseda}, Michael V. and {Matthee}, Jorryt and {Miller}, Tim B. and {Naidu}, Rohan P. and {Nanayakkara}, Themiya and {Oesch}, Pascal A. and {Pan}, Richard and {Papovich}, Casey and {Price}, Sedona H. and {van Dokkum}, Pieter and {Weaver}, John R. and {Whitaker}, Katherine E. and {Zitrin}, Adi},
  title         = {{UNCOVER} Spectroscopy Confirms the Surprising Ubiquity of Active Galactic Nuclei in Red Sources at $z > 5$},
  journal       = {\apj},
  year          = 2024,
  volume        = {964},
  number        = {1},
  pages         = {39},
  doi           = {10.3847/1538-4357/ad1e5f},
  eprint        = {2309.05714},
  archivePrefix = {arXiv},
  primaryClass  = {astro-ph.GA}
}

@article{Kocevski2025,
  author        = {{Kocevski}, Dale D. and {Finkelstein}, Steven L. and {Barro}, Guillermo and {Taylor}, Anthony J. and {Calabr{\`o}}, Antonello and {Laloux}, Brivael and {Buchner}, Johannes and {Trump}, Jonathan R. and {Leung}, Gene C.~K. and {Yang}, Guang and {Dickinson}, Mark and {P{\'e}rez-Gonz{\'a}lez}, Pablo G. and {Pacucci}, Fabio and {Inayoshi}, Kohei and {Somerville}, Rachel S. and {McGrath}, Elizabeth J. and {Akins}, Hollis B. and {Bagley}, Micaela B. and {Bowler}, Rebecca A.~A. and {Bisigello}, Laura and {Carnall}, Adam and {Casey}, Caitlin M. and {Cheng}, Yingjie and {Cleri}, Nikko J. and {Costantin}, Luca and {Cullen}, Fergus and {Davis}, Kelcey and {Donnan}, Callum T. and {Dunlop}, James S. and {Ellis}, Richard S. and {Ferguson}, Henry C. and {Fujimoto}, Seiji and {Fontana}, Adriano and {Giavalisco}, Mauro and {Grazian}, Andrea and {Grogin}, Norman A. and {Hathi}, Nimish P. and {Hirschmann}, Michaela and {Huertas-Company}, Marc and {Holwerda}, Benne W. and {Illingworth}, Garth and {Juneau}, St{\'e}phanie and {Kartaltepe}, Jeyhan S. and {Koekemoer}, Anton M. and {Li}, Wenxiu and {Lucas}, Ray A. and {Magee}, Dan and {Mason}, Charlotte and {McLeod}, Derek J. and {McLure}, Ross J. and {Napolitano}, Lorenzo and {Papovich}, Casey and {Pirzkal}, Nor and {Rodighiero}, Giulia and {Santini}, Paola and {Wilkins}, Stephen M. and {Yung}, L.~Y. Aaron},
  title         = {The Rise of Faint, Red Active Galactic Nuclei at $z > 4$: A Sample of Little Red Dots in the {JWST} Extragalactic Legacy Fields},
  journal       = {\apj},
  year          = 2025,
  volume        = {986},
  number        = {2},
  pages         = {126},
  doi           = {10.3847/1538-4357/adbc7d},
  eprint        = {2404.03576},
  archivePrefix = {arXiv},
  primaryClass  = {astro-ph.GA}
}

@article{Taylor2025,
  author        = {{Taylor}, Anthony J. and {Finkelstein}, Steven L. and {Kocevski}, Dale D. and {Jeon}, Junehyoung and {Bromm}, Volker and {Amor{\'\i}n}, Ricardo O. and {Arrabal Haro}, Pablo and {Backhaus}, Bren E. and {Bagley}, Micaela B. and {Ba{\~n}ados}, Eduardo and {Bhatawdekar}, Rachana and {Brooks}, Madisyn and {Calabr{\`o}}, Antonello and {Ch{\'a}vez Ortiz}, {\'O}scar A. and {Cheng}, Yingjie and {Cleri}, Nikko J. and {Cole}, Justin W. and {Davis}, Kelcey and {Dickinson}, Mark and {Donnan}, Callum and {Dunlop}, James S. and {Ellis}, Richard S. and {Fern{\'a}ndez}, Vital and {Fontana}, Adriano and {Fujimoto}, Seiji and {Giavalisco}, Mauro and {Grazian}, Andrea and {Guo}, Jingsong and {Hathi}, Nimish P. and {Holwerda}, Benne W. and {Hirschmann}, Michaela and {Inayoshi}, Kohei and {Kartaltepe}, Jeyhan S. and {Khusanova}, Yana and {Koekemoer}, Anton M. and {Kokorev}, Vasily and {Larson}, Rebecca L. and {Leung}, Gene C.~K. and {Lucas}, Ray A. and {McLeod}, Derek J. and {Napolitano}, Lorenzo and {Onoue}, Masafusa and {Pacucci}, Fabio and {Papovich}, Casey and {P{\'e}rez-Gonz{\'a}lez}, Pablo G. and {Pirzkal}, Nor and {Somerville}, Rachel S. and {Trump}, Jonathan R. and {Wilkins}, Stephen M. and {Yung}, L.~Y. Aaron and {Zhang}, Haowen},
  title         = {Broad-line {AGNs} at $3.5 < z < 6$: The Black Hole Mass Function and a Connection with Little Red Dots},
  journal       = {\apj},
  year          = 2025,
  volume        = {986},
  number        = {2},
  pages         = {165},
  doi           = {10.3847/1538-4357/add15b},
  eprint        = {2409.06772},
  archivePrefix = {arXiv},
  primaryClass  = {astro-ph.GA}
}

@article{Hviding2025,
  author        = {{Hviding}, Raphael E. and {de Graaff}, Anna and {Miller}, Tim B. and {Setton}, David J. and {Greene}, Jenny E. and {Labb{\'e}}, Ivo and {Brammer}, Gabriel and {Bezanson}, Rachel and {Boogaard}, Leindert A. and {Cleri}, Nikko J. and {Leja}, Joel and {Maseda}, Michael V. and {McConachie}, Ian and {Matthee}, Jorryt and {Naidu}, Rohan P. and {Oesch}, Pascal A. and {Wang}, Bingjie and {Whitaker}, Katherine E. and {Williams}, Christina C.},
  title         = {{RUBIES}: A spectroscopic census of little red dots: All point sources with v-shaped continua have broad lines},
  journal       = {\aap},
  year          = 2025,
  volume        = {702},
  pages         = {A57},
  doi           = {10.1051/0004-6361/202555816},
  eprint        = {2506.05459},
  archivePrefix = {arXiv},
  primaryClass  = {astro-ph.GA}
}

@article{BaccusXu2026,
  author        = {{Baccus}, Caroline J. and {Xu}, Xinfeng},
  title         = {A Comprehensive {JWST/NIRSpec} Census of Broad-line Active Galactic Nuclei: Faint, Low-mass Accretors and a Population of Little Red Dots},
  journal       = {\apj},
  year          = 2026,
  volume        = {1006},
  number        = {2},
  pages         = {165},
  doi           = {10.3847/1538-4357/ae7de7},
  eprint        = {2512.03281},
  archivePrefix = {arXiv},
  primaryClass  = {astro-ph.GA}
}

@article{Setton2025Vshape,
  author        = {{Setton}, David J. and {Greene}, Jenny E. and {de Graaff}, Anna and {Ma}, Yilun and {Leja}, Joel and {Matthee}, Jorryt and {Bezanson}, Rachel and {Boogaard}, Leindert A. and {Cleri}, Nikko J. and {Katz}, Harley and {Labb{\'e}}, Ivo and {Maseda}, Michael V. and {McConachie}, Ian and {Miller}, Tim B. and {Price}, Sedona H. and {Suess}, Katherine A. and {van Dokkum}, Pieter and {Wang}, Bingjie and {Weibel}, Andrea and {Whitaker}, Katherine E. and {Williams}, Christina C.},
  title         = {Little Red Dots at an Inflection Point: Ubiquitous ``{V}-Shaped'' Turnover Consistently Occurs at the Balmer Limit},
  journal       = {\apj},
  year          = 2025,
  volume        = {995},
  number        = {1},
  pages         = {118},
  doi           = {10.3847/1538-4357/ae1500},
  eprint        = {2411.03424},
  archivePrefix = {arXiv},
  primaryClass  = {astro-ph.GA}
}

@article{InayoshiMaiolino2025,
  author        = {{Inayoshi}, Kohei and {Maiolino}, Roberto},
  title         = {Extremely Dense Gas around Little Red Dots and High-redshift Active Galactic Nuclei: A Non-stellar Origin of the Balmer Break and Absorption Features},
  journal       = {\apjl},
  year          = 2025,
  volume        = {980},
  number        = {2},
  pages         = {L27},
  doi           = {10.3847/2041-8213/adaebd},
  eprint        = {2409.07805},
  archivePrefix = {arXiv},
  primaryClass  = {astro-ph.GA}
}

@article{deGraaffRuby2025,
  author        = {{de Graaff}, Anna and {Rix}, Hans-Walter and {Naidu}, Rohan P. and {Labb{\'e}}, Ivo and {Wang}, Bingjie and {Leja}, Joel and {Matthee}, Jorryt and {Katz}, Harley and {Greene}, Jenny E. and {Hviding}, Raphael E. and {Baggen}, Josephine and {Bezanson}, Rachel and {Boogaard}, Leindert A. and {Brammer}, Gabriel and {Dayal}, Pratika and {van Dokkum}, Pieter and {Goulding}, Andy D. and {Hirschmann}, Michaela and {Maseda}, Michael V. and {McConachie}, Ian and {Miller}, Tim B. and {Nelson}, Erica and {Oesch}, Pascal A. and {Setton}, David J. and {Shivaei}, Irene and {Weibel}, Andrea and {Whitaker}, Katherine E. and {Williams}, Christina C.},
  title         = {A remarkable ruby: Absorption in dense gas, rather than evolved stars, drives the extreme {Balmer} break of a little red dot at $z=3.5$},
  journal       = {\aap},
  year          = 2025,
  volume        = {701},
  pages         = {A168},
  doi           = {10.1051/0004-6361/202554681},
  eprint        = {2503.16600},
  archivePrefix = {arXiv},
  primaryClass  = {astro-ph.GA}
}

@article{NaiduBHStar2026,
  author        = {{Naidu}, Rohan P. and {Matthee}, Jorryt and {Katz}, Harley and {de Graaff}, Anna and {Oesch}, Pascal and {Smith}, Aaron and {Greene}, Jenny E. and {Brammer}, Gabriel and {Weibel}, Andrea and {Hviding}, Raphael and {Chisholm}, John and {Labb{\'e}}, Ivo and {Simcoe}, Robert A. and {Witten}, Callum and {Atek}, Hakim and {Baggen}, Josephine F. W. and {Belli}, Sirio and {Bezanson}, Rachel and {Boogaard}, Leindert A. and {Bose}, Sownak and {Covelo-Paz}, Alba and {Dayal}, Pratika and {Fudamoto}, Yoshinobu and {Furtak}, Lukas J. and {Giovinazzo}, Emma and {Goulding}, Andy and {Gronke}, Max and {Heintz}, Kasper E. and {Hirschmann}, Michaela and {Illingworth}, Garth and {Inoue}, Akio K. and {Johnson}, Benjamin D. and {Leja}, Joel and {Leonova}, Ecaterina and {McConachie}, Ian and {Maseda}, Michael V. and {Natarajan}, Priyamvada and {Nelson}, Erica and {Setton}, David J. and {Shivaei}, Irene and {Sobral}, David and {Stefanon}, Mauro and {Tacchella}, Sandro and {Toft}, Sune and {Torralba}, Alberto and {van Dokkum}, Pieter and {van der Wel}, Arjen and {Volonteri}, Marta and {Walter}, Fabian and {Wang}, Bingjie and {Watson}, Darach},
  title         = {A gas-enshrouded and gas-reddened black hole at cosmic dawn},
  journal       = {\nat},
  year          = 2026,
  volume        = {656},
  pages         = {329--333},
  doi           = {10.1038/s41586-026-10846-4},
  eprint        = {2503.16596},
  archivePrefix = {arXiv},
  primaryClass  = {astro-ph.GA}
}

@article{DEugenio2026,
  author        = {{D'Eugenio}, Francesco and {Maiolino}, Roberto and {Perna}, Michele and {{\"U}bler}, Hannah and {Ji}, Xihan and {McClymont}, William and {Koudmani}, Sophie and {Sijacki}, Debora and {Juod{\v{z}}balis}, Ignas and {Scholtz}, Jan and {Bennett}, Jake S. and {Bunker}, Andrew J. and {Carniani}, Stefano and {Charlot}, St{\'e}phane and {Cresci}, Giovanni and {Curtis-Lake}, Emma and {Dalla Bont{\`a}}, Elena and {Inayoshi}, Kohei and {Jones}, Gareth C. and {Lyu}, Jianwei and {Marconi}, Alessandro and {Mazzolari}, Giovanni and {Nelson}, Erica J. and {Parlanti}, Eleonora and {Robertson}, Brant E. and {Schneider}, Raffaella and {Simmonds}, Charlotte and {Tacchella}, Sandro and {Venturi}, Giacomo and {Willott}, Chris and {Witstok}, Joris and {Witten}, Callum},
  title         = {{BlackTHUNDER} strikes twice: {Balmer}-line absorption in an overmassive Little Red Dot at $z = 7.04$},
  journal       = {\mnras},
  year          = 2026,
  volume        = {547},
  number        = {4},
  eid           = {stag401},
  pages         = {stag401},
  doi           = {10.1093/mnras/stag401},
  eprint        = {2503.11752},
  archivePrefix = {arXiv},
  primaryClass  = {astro-ph.GA}
}

@article{deGraaff2026,
  author        = {{de Graaff}, Anna and {Hviding}, Raphael E. and {Naidu}, Rohan P. and {Greene}, Jenny E. and {Miller}, Tim B. and {Leja}, Joel and {Matthee}, Jorryt and {Brammer}, Gabriel and {Katz}, Harley and {Bezanson}, Rachel and {Boogaard}, Leindert A. and {Bose}, Sownak and {Chisholm}, John and {Cleri}, Nikko J. and {Dayal}, Pratika and {Feldmann}, Robert and {Fudamoto}, Yoshinobu and {Fujimoto}, Seiji and {Furtak}, Lukas J. and {Glazebrook}, Karl and {Gottumukkala}, Rashmi and {Heintz}, Kasper E. and {Kokorev}, Vasily and {Labb{\'e}}, Ivo and {Maseda}, Michael V. and {McConachie}, Ian and {Nanayakkara}, Themiya and {Nelson}, Erica and {Nowaczyk}, Przemys{\l}aw and {Oesch}, Pascal A. and {Rix}, Hans-Walter and {Setton}, David J. and {Torralba}, Alberto and {Walter}, Fabian and {Wang}, Bingjie and {Weibel}, Andrea and {van der Wel}, Arjen},
  title         = {Little red dots host black hole stars: a unified family of gas-reddened {AGN} revealed by {JWST/NIRSpec} spectroscopy},
  journal       = {\mnras},
  year          = 2026,
  volume        = {551},
  number        = {3},
  eid           = {stag1567},
  pages         = {stag1567},
  doi           = {10.1093/mnras/stag1567},
  eprint        = {2511.21820},
  archivePrefix = {arXiv},
  primaryClass  = {astro-ph.GA}
}

@article{InayoshiHo,
  author        = {{Inayoshi}, Kohei and {Ho}, Luis C.},
  title         = {A critical evaluation of the physical nature of the little red dots},
  journal       = {Nature Astronomy},
  year          = 2026,
  volume        = {10},
  number        = {9},
  pages         = {1261--1272},
  doi           = {10.1038/s41550-026-02934-2}
}

@article{Rusakov2026,
  author        = {{Rusakov}, V. and {Watson}, D. and {Nikopoulos}, G.~P. and {Brammer}, G. and {Gottumukkala}, R. and {Harvey}, T. and {Heintz}, K.~E. and {Damgaard}, R. and {Sim}, S.~A. and {Sneppen}, A. and {Vijayan}, A.~P. and {Adams}, N. and {Austin}, D. and {Conselice}, C.~J. and {Goolsby}, C.~M. and {Toft}, S. and {Witstok}, J.},
  title         = {Little red dots as young supermassive black holes in dense ionized cocoons},
  journal       = {\nat},
  year          = 2026,
  volume        = {649},
  number        = {8097},
  pages         = {574--579},
  doi           = {10.1038/s41586-025-09900-4},
  eprint        = {2503.16595},
  archivePrefix = {arXiv},
  primaryClass  = {astro-ph.GA}
}

@article{Chang2025,
  author        = {{Chang}, Seok-Jun and {Gronke}, Max and {Matthee}, Jorryt and {Mason}, Charlotte},
  title         = {Impact of resonance, {Raman}, and {Thomson} scattering on hydrogen line formation in Little Red Dots},
  journal       = {\mnras},
  year          = 2026,
  volume        = {545},
  number        = {4},
  eid           = {staf2131},
  pages         = {staf2131},
  doi           = {10.1093/mnras/staf2131},
  eprint        = {2508.08768},
  archivePrefix = {arXiv},
  primaryClass  = {astro-ph.GA}
}

@article{Brazzini2025,
  author        = {{Brazzini}, Matilde and {D'Eugenio}, Francesco and {Maiolino}, Roberto and {Juod{\v{z}}balis}, Ignas and {Ji}, Xihan and {Scholtz}, Jan and {Chang}, Seok-Jun},
  title         = {Ruling out dominant electron scattering in Little Red Dots' {Rosetta Stone} using multiple hydrogen lines},
  journal       = {\mnras},
  year          = 2025,
  volume        = {544},
  pages         = {L167--L173},
  doi           = {10.1093/mnrasl/slaf116},
  eprint        = {2507.08929},
  archivePrefix = {arXiv},
  primaryClass  = {astro-ph.GA}
}

@article{Madau2026b,
  author        = {{Madau}, Piero and {Maiolino}, Roberto and {Scholtz}, Jan and {D'Eugenio}, Francesco},
  title         = {Wings of little dots: Exponential broad lines from a stratified broad-line region},
  journal       = {\aap},
  year          = 2026,
  volume        = {714},
  eid           = {A186},
  pages         = {A186},
  doi           = {10.1051/0004-6361/202660251},
  eprint        = {2604.04216},
  archivePrefix = {arXiv},
  primaryClass  = {astro-ph.GA}
}

@article{Scholtz2026,
  author        = {{Scholtz}, J. and {D'Eugenio}, F. and {Maiolino}, R. and {Brazzini}, M. and {{\"U}bler}, H. and {Ji}, X. and {Perna}, M. and {Sun}, F. and {Brocchi}, G. and {Carniani}, S. and {Cresci}, G. and {Ivey}, L.~R. and {Juod{\v{z}}balis}, I. and {Marconi}, A. and {Mazzolari}, G. and {Risaliti}, G. and {Trefoloni}, B.},
  title         = {Little Red and Blue Dots: simply stratified Broad Line Regions},
  journal       = {arXiv e-prints},
  year          = 2026,
  eid           = {arXiv:2603.22277},
  pages         = {arXiv:2603.22277},
  doi           = {10.48550/arXiv.2603.22277},
  eprint        = {2603.22277},
  archivePrefix = {arXiv},
  primaryClass  = {astro-ph.GA}
}

@article{Matthee2026,
  author        = {{Matthee}, Jorryt and {Torralba}, Alberto and {Pezzulli}, Gabriele and {Naidu}, Rohan P. and {Chisholm}, John and {Mascia}, Sara and {Greene}, Jenny E. and {Ishikawa}, Yuzo and {Gronke}, Max and {Wuyts}, Stijn and {Bordoloi}, Rongmon and {Brammer}, Gabriel and {Chang}, Seok-Jun and {Eilers}, Anna-Christina and {de Graaff}, Anna and {Hviding}, Raphael E. and {Iani}, Edoardo and {Illingworth}, Garth and {Kashino}, Daichi and {Labb{\'e}}, Ivo and {Ma}, Yilun and {Maseda}, Michael V. and {Meyer}, Romain and {Nelson}, Erica and {Oesch}, Pascal and {Xiao}, Mengyuan},
  title         = {The Engine and its Flows: Little Red Dot spectra are shaped by the column densities of their gas envelopes},
  journal       = {arXiv e-prints},
  year          = 2026,
  eid           = {arXiv:2603.17667},
  pages         = {arXiv:2603.17667},
  doi           = {10.48550/arXiv.2603.17667},
  eprint        = {2603.17667},
  archivePrefix = {arXiv},
  primaryClass  = {astro-ph.GA}
}

@article{Shen2011,
  author        = {{Shen}, Yue and {Richards}, Gordon T. and {Strauss}, Michael A. and {Hall}, Patrick B. and {Schneider}, Donald P. and {Snedden}, Stephanie and {Bizyaev}, Dmitry and {Brewington}, Howard and {Malanushenko}, Viktor and {Malanushenko}, Elena and {Oravetz}, Dan and {Pan}, Kaike and {Simmons}, Audrey},
  title         = {A Catalog of Quasar Properties from {Sloan Digital Sky Survey} Data Release 7},
  journal       = {\apjs},
  year          = 2011,
  volume        = {194},
  number        = {2},
  pages         = {45},
  doi           = {10.1088/0067-0049/194/2/45}
}

@article{GreeneHo2005,
  author        = {{Greene}, Jenny E. and {Ho}, Luis C.},
  title         = {Estimating Black Hole Masses in Active Galaxies Using the {H$\alpha$} Emission Line},
  journal       = {\apj},
  year          = 2005,
  volume        = {630},
  number        = {1},
  pages         = {122--129},
  doi           = {10.1086/431897}
}

@article{Peterson2004,
  author        = {{Peterson}, Bradley M. and {Ferrarese}, Laura and {Gilbert}, Karoline M. and {Kaspi}, Shai and {Malkan}, Matthew A. and {Maoz}, Dan and {Merritt}, David and {Netzer}, Hagai and {Onken}, Christopher A. and {Pogge}, Richard W. and {Vestergaard}, Marianne and {Wandel}, Amri},
  title         = {Central Masses and Broad-Line Region Sizes of Active Galactic Nuclei. {II.} A Homogeneous Analysis of a Large Reverberation-Mapping Database},
  journal       = {\apj},
  year          = 2004,
  volume        = {613},
  number        = {2},
  pages         = {682--699},
  doi           = {10.1086/423269}
}

@article{Collin2006,
  author        = {{Collin}, S. and {Kawaguchi}, T. and {Peterson}, B.~M. and {Vestergaard}, M.},
  title         = {Systematic effects in measurement of black hole masses by emission-line reverberation of active galactic nuclei: {Eddington} ratio and inclination},
  journal       = {\aap},
  year          = 2006,
  volume        = {456},
  number        = {1},
  pages         = {75--90},
  doi           = {10.1051/0004-6361:20064878}
}

@article{Bentz2013,
  author        = {{Bentz}, Misty C. and {Denney}, Kelly D. and {Grier}, Catherine J. and {Barth}, Aaron J. and {Peterson}, Bradley M. and {Vestergaard}, Marianne and {Bennert}, Vardha N. and {Canalizo}, Gabriela and {De Rosa}, Gisella and {Filippenko}, Alexei V. and {Gates}, Elinor L. and {Greene}, Jenny E. and {Li}, Weidong and {Malkan}, Matthew A. and {Pogge}, Richard W. and {Stern}, Daniel and {Treu}, Tommaso and {Woo}, Jong-Hak},
  title         = {The Low-luminosity End of the Radius--Luminosity Relationship for Active Galactic Nuclei},
  journal       = {\apj},
  year          = 2013,
  volume        = {767},
  number        = {2},
  pages         = {149},
  doi           = {10.1088/0004-637X/767/2/149}
}

@article{Fine2008,
  author        = {{Fine}, S. and {Croom}, S.~M. and {Hopkins}, P.~F. and {Hernquist}, L. and {Bland-Hawthorn}, J. and {Colless}, M. and {Hall}, P.~B. and {Miller}, L. and {Myers}, A.~D. and {Nichol}, R. and {Pimbblet}, K.~A. and {Ross}, N.~P. and {Schneider}, D.~P. and {Shanks}, T. and {Sharp}, R.~G.},
  title         = {Constraining the quasar population with the broad-line width distribution},
  journal       = {\mnras},
  year          = 2008,
  volume        = {390},
  number        = {4},
  pages         = {1413--1429},
  doi           = {10.1111/j.1365-2966.2008.13691.x}
}

@article{Croom2011,
  author        = {{Croom}, Scott M.},
  title         = {Do Quasar Broad-line Velocity Widths Add Any Information to Virial Black Hole Mass Estimates?},
  journal       = {\apj},
  year          = 2011,
  volume        = {736},
  number        = {2},
  pages         = {161},
  doi           = {10.1088/0004-637X/736/2/161}
}

@article{Assef2012,
  author        = {{Assef}, R.~J. and {Frank}, S. and {Grier}, C.~J. and {Kochanek}, C.~S. and {Denney}, K.~D. and {Peterson}, B.~M.},
  title         = {The Importance of Broad Emission Line Widths in Single-epoch Black Hole Mass Estimates},
  journal       = {\apjl},
  year          = 2012,
  volume        = {753},
  number        = {1},
  pages         = {L2},
  doi           = {10.1088/2041-8205/753/1/L2}
}

@article{Laor2006,
  author        = {{Laor}, Ari},
  title         = {Evidence for Line Broadening by Electron Scattering in the Broad-Line Region of {NGC 4395}},
  journal       = {\apj},
  year          = 2006,
  volume        = {643},
  number        = {1},
  pages         = {112--119},
  doi           = {10.1086/502798}
}

@article{Jakobsen2022,
  author        = {{Jakobsen}, P. and {Ferruit}, P. and {Alves de Oliveira}, C. and {Arribas}, S. and {Bagnasco}, G. and {Barho}, R. and {Beck}, T.~L. and {Birkmann}, S. and {B{\"o}ker}, T. and {Bunker}, A.~J. and {Charlot}, S. and {de Jong}, P. and {de Marchi}, G. and {Ehrenwinkler}, R. and {Falcolini}, M. and {Fels}, R. and {Franx}, M. and {Franz}, D. and {Funke}, M. and {Giardino}, G. and {Gnata}, X. and {Holota}, W. and {Honnen}, K. and {Jensen}, P.~L. and {Jentsch}, M. and {Johnson}, T. and {Jollet}, D. and {Karl}, H. and {Kling}, G. and {K{\"o}hler}, J. and {Kolm}, M.-G. and {Kumari}, N. and {Lander}, M.~E. and {Lemke}, R. and {L{\'o}pez-Caniego}, M. and {L{\"u}tzgendorf}, N. and {Maiolino}, R. and {Manjavacas}, E. and {Marston}, A. and {Maschmann}, M. and {Maurer}, R. and {Messerschmidt}, B. and {Moseley}, S.~H. and {Mosner}, P. and {Mott}, D.~B. and {Muzerolle}, J. and {Pirzkal}, N. and {Pittet}, J.-F. and {Plitzke}, A. and {Posselt}, W. and {Rapp}, B. and {Rauscher}, B.~J. and {Rawle}, T. and {Rix}, H.-W. and {R{\"o}del}, A. and {Rumler}, P. and {Sabbi}, E. and {Salvignol}, J.-C. and {Schmid}, T. and {Sirianni}, M. and {Smith}, C. and {Strada}, P. and {te Plate}, M. and {Valenti}, J. and {Wettemann}, T. and {Wiehe}, T. and {Wiesmayer}, M. and {Willott}, C.~J. and {Wright}, R. and {Zeidler}, P. and {Zincke}, C.},
  title         = {The {Near-Infrared Spectrograph (NIRSpec)} on the {James Webb Space Telescope. I.} Overview of the instrument and its capabilities},
  journal       = {\aap},
  year          = 2022,
  volume        = {661},
  eid           = {A80},
  pages         = {A80},
  doi           = {10.1051/0004-6361/202142663}
}

@article{deGraaffLSF2024,
  author        = {{de Graaff}, Anna and {Rix}, Hans-Walter and {Carniani}, Stefano and {Suess}, Katherine A. and {Charlot}, St{\'e}phane and {Curtis-Lake}, Emma and {Arribas}, Santiago and {Baker}, William M. and {Boyett}, Kristan and {Bunker}, Andrew J. and {Cameron}, Alex J. and {Chevallard}, Jacopo and {Curti}, Mirko and {Eisenstein}, Daniel J. and {Franx}, Marijn and {Hainline}, Kevin and {Hausen}, Ryan and {Ji}, Zhiyuan and {Johnson}, Benjamin D. and {Jones}, Gareth C. and {Maiolino}, Roberto and {Maseda}, Michael V. and {Nelson}, Erica and {Parlanti}, Eleonora and {Rawle}, Tim and {Robertson}, Brant and {Tacchella}, Sandro and {{\"U}bler}, Hannah and {Williams}, Christina C. and {Willmer}, Christopher N.~A. and {Willott}, Chris},
  title         = {Ionised gas kinematics and dynamical masses of $z \gtrsim 6$ galaxies from {JADES/NIRSpec} high-resolution spectroscopy},
  journal       = {\aap},
  year          = 2024,
  volume        = {684},
  eid           = {A87},
  pages         = {A87},
  doi           = {10.1051/0004-6361/202347755}
}

@article{DJAdeGraaff2024,
  author        = {{de Graaff}, Anna and {Brammer}, Gabriel and {Weibel}, Andrea and {Lewis}, Zach and {Maseda}, Michael V. and {Oesch}, Pascal A. and {Bezanson}, Rachel and {Boogaard}, Leindert A. and {Cleri}, Nikko J. and {Cooper}, Olivia R. and {Gottumukkala}, Rashmi and {Greene}, Jenny E. and {Hirschmann}, Michaela and {Hviding}, Raphael E. and {Katz}, Harley and {Labb{\'e}}, Ivo and {Leja}, Joel and {Matthee}, Jorryt and {McConachie}, Ian and {Miller}, Tim B. and {Naidu}, Rohan P. and {Price}, Sedona H. and {Rix}, Hans-Walter and {Setton}, David J. and {Suess}, Katherine A. and {Wang}, Bingjie and {Whitaker}, Katherine E. and {Williams}, Christina C.},
  title         = {{RUBIES}: A complete census of the bright and red distant {Universe} with {JWST/NIRSpec}},
  journal       = {\aap},
  year          = 2025,
  volume        = {697},
  eid           = {A189},
  pages         = {A189},
  doi           = {10.1051/0004-6361/202452186}
}

@misc{msaexp,
  author        = {{Brammer}, Gabriel},
  title         = {msaexp: {NIRSpec} analysis tools, v0.6.17},
  year          = 2023,
  howpublished  = {Zenodo},
  doi           = {10.5281/zenodo.7299500}
}

@article{Planck2020,
  author        = {{Planck Collaboration} and {Aghanim}, N. and {Akrami}, Y. and {Ashdown}, M. and {Aumont}, J. and {Baccigalupi}, C. and {Ballardini}, M. and others},
  title         = {{Planck} 2018 results. {VI.} Cosmological parameters},
  journal       = {\aap},
  year          = 2020,
  volume        = {641},
  eid           = {A6},
  pages         = {A6},
  doi           = {10.1051/0004-6361/201833910}
}

@article{EuclidAussel2026,
  author        = {{Euclid Collaboration} and {Aussel}, H. and {Cuillandre}, J.-C. and {Mellier}, Y. and {Altieri}, B. and {Amara}, A. and others},
  title         = {{Euclid} First Data Release 1 ({DR1}) -- Data release overview},
  journal       = {arXiv e-prints},
  year          = 2026,
  eid           = {arXiv:2611.16510},
  pages         = {arXiv:2611.16510},
  doi           = {10.48550/arXiv.2611.16510},
  eprint        = {2611.16510},
  archivePrefix = {arXiv},
  primaryClass  = {astro-ph.CO}
}

@article{EuclidLeBrun2026,
  author        = {{Euclid Collaboration} and {Le Brun}, V. and {Moresco}, M. and {Bethermin}, M. and {Franzetti}, P. and {Fumana}, M. and others},
  title         = {{Euclid} Data Release 1 ({DR1}): Performance of the redshift measurement and spectroscopic analysis},
  journal       = {arXiv e-prints},
  year          = 2026,
  eid           = {arXiv:2611.17960},
  pages         = {arXiv:2611.17960},
  doi           = {10.48550/arXiv.2611.17960},
  eprint        = {2611.17960},
  archivePrefix = {arXiv},
  primaryClass  = {astro-ph.CO}
}

@article{EuclidFumana2026,
  author        = {{Euclid Collaboration} and {Fumana}, M. and {Copin}, Y. and {Appleton}, P.~N. and {Jacobson}, J. and {Zoubian}, J. and others},
  title         = {{Euclid} data release 1 ({DR1}) -- {SIR} spectroscopic calibration},
  journal       = {arXiv e-prints},
  year          = 2026,
  eid           = {arXiv:2611.17305},
  pages         = {arXiv:2611.17305},
  doi           = {10.48550/arXiv.2611.17305},
  eprint        = {2611.17305},
  archivePrefix = {arXiv},
  primaryClass  = {astro-ph.IM}
}

@article{EuclidBisigello2025,
  author        = {{Euclid Collaboration} and {Bisigello}, L. and {Rodighiero}, G. and {Fotopoulou}, S. and {Bianchini}, F. and {Bolzonella}, M. and others},
  title         = {{Euclid} Quick Data Release ({Q1}). {XXIV.} Extending the quest for little red dots to $z < 4$},
  journal       = {\aap},
  year          = 2026,
  volume        = {711},
  eid           = {A24},
  pages         = {A24},
  doi           = {10.1051/0004-6361/202554537},
  eprint        = {2503.15323},
  archivePrefix = {arXiv},
  primaryClass  = {astro-ph.GA}
}

@book{OsterbrockFerland2006,
  author        = {{Osterbrock}, Donald E. and {Ferland}, Gary J.},
  title         = {Astrophysics of Gaseous Nebulae and Active Galactic Nuclei},
  edition       = {2nd},
  publisher     = {University Science Books},
  address       = {Sausalito, CA},
  year          = 2006
}

@article{harris2020array,
  author        = {Harris, Charles R. and Millman, K. Jarrod and {van der Walt}, St{\'e}fan J. and Gommers, Ralf and Virtanen, Pauli and Cournapeau, David and Wieser, Eric and Taylor, Julian and Berg, Sebastian and Smith, Nathaniel J. and Kern, Robert and Picus, Matti and Hoyer, Stephan and {van Kerkwijk}, Marten H. and Brett, Matthew and Haldane, Allan and {Fern{\'a}ndez del R{\'\i}o}, Jaime and Wiebe, Mark and Peterson, Pearu and {G{\'e}rard-Marchant}, Pierre and Sheppard, Kevin and Reddy, Tyler and Weckesser, Warren and Abbasi, Hameer and Gohlke, Christoph and Oliphant, Travis E.},
  title         = {Array programming with {NumPy}},
  journal       = {\nat},
  year          = 2020,
  volume        = {585},
  number        = {7825},
  pages         = {357--362},
  doi           = {10.1038/s41586-020-2649-2}
}

@article{2020SciPy-NMeth,
  author        = {Virtanen, Pauli and Gommers, Ralf and Oliphant, Travis E. and Haberland, Matt and Reddy, Tyler and Cournapeau, David and Burovski, Evgeni and Peterson, Pearu and Weckesser, Warren and Bright, Jonathan and {van der Walt}, St{\'e}fan J. and Brett, Matthew and Wilson, Joshua and Millman, K. Jarrod and Mayorov, Nikolay and Nelson, Andrew R. J. and Jones, Eric and Kern, Robert and Larson, Eric and Carey, C. J. and Polat, {\.I}lhan and Feng, Yu and Moore, Eric W. and {VanderPlas}, Jake and Laxalde, Denis and Perktold, Josef and Cimrman, Robert and Henriksen, Ian and Quintero, E. A. and Harris, Charles R. and Archibald, Anne M. and Ribeiro, Ant{\^o}nio H. and Pedregosa, Fabian and {van Mulbregt}, Paul and {SciPy 1.0 Contributors}},
  title         = {{SciPy} 1.0: Fundamental Algorithms for Scientific Computing in {Python}},
  journal       = {Nature Methods},
  year          = 2020,
  volume        = {17},
  pages         = {261--272},
  doi           = {10.1038/s41592-019-0686-2}
}

@article{Hunter:2007,
  author        = {Hunter, J. D.},
  title         = {{Matplotlib}: A {2D} graphics environment},
  journal       = {Computing in Science \& Engineering},
  year          = 2007,
  volume        = {9},
  number        = {3},
  pages         = {90--95},
  doi           = {10.1109/MCSE.2007.55}
}

@article{astropy:2013,
  author        = {{Astropy Collaboration} and {Robitaille}, Thomas P. and {Tollerud}, Erik J. and {Greenfield}, Perry and {Droettboom}, Michael and {Bray}, Erik and {Aldcroft}, Tom and others},
  title         = {{Astropy}: A community {Python} package for astronomy},
  journal       = {\aap},
  year          = 2013,
  volume        = {558},
  eid           = {A33},
  pages         = {A33},
  doi           = {10.1051/0004-6361/201322068}
}

@article{astropy:2018,
  author        = {{Astropy Collaboration} and {Price-Whelan}, A.~M. and {Sip{\H{o}}cz}, B.~M. and {G{\"u}nther}, H.~M. and {Lim}, P.~L. and {Crawford}, S.~M. and {Conseil}, S. and others},
  title         = {The {Astropy} Project: Building an Open-science Project and Status of the v2.0 Core Package},
  journal       = {\aj},
  year          = 2018,
  volume        = {156},
  number        = {3},
  pages         = {123},
  doi           = {10.3847/1538-3881/aabc4f}
}

@article{astropy:2022,
  author        = {{Astropy Collaboration} and {Price-Whelan}, Adrian M. and {Lim}, Pey Lian and {Earl}, Nicholas and {Starkman}, Nathaniel and {Bradley}, Larry and {Shupe}, David L. and others},
  title         = {The {Astropy} Project: Sustaining and Growing a Community-oriented Open-source Project and the Latest Major Release (v5.0) of the Core Package},
  journal       = {\apj},
  year          = 2022,
  volume        = {935},
  number        = {2},
  pages         = {167},
  doi           = {10.3847/1538-4357/ac7c74}
}

@article{Yue2024,
  author = {Yue, Minghao and Eilers, Anna-Christina and Ananna, Tonima Tasnim and Panagiotou, Christos and Kara, Erin and Miyaji, Takamitsu},
  title = {Stacking X-Ray Observations of {Little Red Dots}: Implications for Their Active Galactic Nucleus Properties},
  year = {2024},
  journal = {\apjl},
  volume = {974},
  pages = {L26},
  doi = {10.3847/2041-8213/ad7eba}
}
\bibliographystyle{aasjournal}

\end{document}